\documentclass[preprint2]{aastex}

\input psfig
\input epsf

\newcommand{\esoq}{ESO215--G?009}
\newcommand{\hi}{H\,{\sc i}}
\newcommand{\hii}{H\,{\sc ii}}
\newcommand{\mlr}{${\cal M}_{\rm HI}/L_{\rm B}$}
\newcommand{\mtmr}{${\cal M}_{\rm HI}/{\cal M}_{\rm tot}$}
\newcommand{\bmtmr}{${\cal M}_{\rm bary}/{\cal M}_{\rm tot}$}
\newcommand{\tmlr}{${\cal M}_{\rm tot}/L_{\rm B}$}
\newcommand{\mls}{~${\cal M}_{\sun}/L_{\sun,{\rm B}}$}
\newcommand{\mB}{$m_{\rm B}$}
\newcommand{\mR}{$m_{\rm R}$}
\newcommand{\MB}{$M_{\rm B}$}
\newcommand{\AB}{$A_{\rm B}$}
\newcommand{\AG}{$A_{\rm G}$}
\newcommand{\LB}{$L_{\rm B}$}
\newcommand{\Lsun}{~$L_{\sun,{\rm B}}$}
\newcommand{\LLsun}{$L_{\sun,{\rm B}}$}
\newcommand{\Msun}{~${\cal M}_{\sun}$}
\newcommand{\MMsun}{${\cal M}_{\sun}$}
\newcommand{\FHI}{$F_{\rm HI}$}
\newcommand{\MHI}{${\cal M}_{\rm HI}$}
\newcommand{\Mbary}{${\cal M}_{\rm bary}$}
\newcommand{\Mtot}{${\cal M}_{\rm tot}$}
\newcommand{\kms}{~km\,s$^{-1}$}
\newcommand{\jks}{~Jy~km\,s$^{-1}$}
\newcommand{\jjks}{Jy~km\,s$^{-1}$}
\newcommand{\vmax}{$v_{\rm max}$}

\slugcomment{}

\shorttitle{The Dwarf Irregular Galaxy ESO215--G?009}
\shortauthors{Warren, Jerjen \& Koribalski}

\begin{document}
\title{\objectname[]{ESO215--G?009}:  An Extreme \hi-Rich Dwarf Irregular Galaxy}
\author{Bradley E. Warren\altaffilmark{1} and Helmut Jerjen}
\affil{Research School of Astronomy and Astrophysics, Australian National 
       University, Mount Stromlo Observatory, Cotter Road, Weston ACT 2611, 
       Australia}
\email{bewarren@mso.anu.edu.au, jerjen@mso.anu.edu.au\\}
\and
\author{B\"arbel S. Koribalski}
\affil{Australia Telescope National Facility, CSIRO, PO Box 76, Epping NSW 1710, Australia}
\email{Baerbel.Koribalski@csiro.au}

\altaffiltext{1}{Affiliated with the Australia Telescope National Facility, CSIRO.}

\begin{abstract}
We present deep {\em BVRI} band images and \hi\ line observations of the nearby, low surface brightness galaxy \esoq\ which were obtained with the ANU 2.3-m Telescope and the Australia Telescope Compact Array, respectively. \esoq\ was selected from the HIPASS {\it Bright Galaxy Catalog} because it has the second highest \hi\ mass-to-light ratio of the galaxies with measured {\em B} band apparent magnitudes.  We find that it is an isolated dwarf irregular galaxy with an old stellar population.  We place an upper limit on the current star formation rate of $\sim 2.5 \times 10^{-3}$\Msun\,yr$^{-1}$.  The extended \hi\ disk shows regular rotation ($v_{rot} = 51 \pm 8$\kms), and at a column density of $\sim 5.0 \times 10^{19}$ atoms\,cm$^{-2}$ can be traced out to over six times the Holmberg radius of the stellar component (radius at $\mu_{B} = 26.6$ mag arcsec$^{-2}$).  After foreground star subtraction, we measure a {\em B} band apparent magnitude of $16.13 \pm 0.07$ mag within a radius of 80\arcsec.  The \hi\ flux density is $122 \pm 4$\jks\ within a radius of 370\arcsec. Given a Galactic extinction of \AB\ = $0.95 \pm 0.15$ mag, we derive an \hi\ mass-to-light ratio of $22 \pm 4$\mls\ for \esoq.  To our knowledge this is the highest \mlr\ ratio for a galaxy to be confirmed by accurate measurement to date.
\end{abstract}


\section{Introduction}
One way to investigate aspects of star formation and galaxy evolution is to study galaxies which appear to have done little of either, those with large quantities of unprocessed neutral hydrogen (\hi) compared to their stellar content. The high quantities of \hi\ gas in some low luminosity galaxies suggest that star formation within these galaxies has been impaired or halted, has lacked stimulation, or has only recently begun.

The \hi\ mass-to-light ratio of a galaxy is a distance independent quantity which compares the \hi\ mass to the luminosity in a particular photometric band. As the {\em B} band magnitude of galaxies has been in the past the most commonly available (from photographic studies), the \hi\ mass-to-light ratio is usually expressed with the {\em B} band luminosity:
\begin{equation}  
\frac{{\cal M}_{\rm HI}}{L_{\rm B}} = 1.5 \times 10^{-7} F_{\rm HI}~ 10^{0.4(m_{\rm B}-A_{\rm B})}~~\frac{{\cal M}_{\sun}}{L_{\sun,{\rm B}}} ,
\label{eqn:mlr}
\end{equation} 
where \MHI\ is the \hi\ mass in \MMsun, \LB\ is the {\em B} band luminosity in \LLsun, \FHI\ is the integrated \hi\ flux density in \jjks, \mB\ is the {\em B} band apparent magnitude, and \AB\ is the {\em B} band Galactic extinction. Extinction correction from the host galaxy is not included (see Section~\ref{sec:dis-uncert}).

For most galaxies \mlr\ is typically less than 1\mls, with late-type galaxies having a larger spread than early-types.  \citet{rob94} performed a statistical study of physical parameters along the Hubble sequence.  For late-type galaxies they found a general trend of lower luminosity, surface brightness and \hi\ mass, and higher \mlr\ and \hi-to-total mass ratio than for early-type spirals.  E.g., for the RC3 local Supercluster sample they find a median \mlr\ of 0.78 \mls\ in type Sm/Im galaxies, with an inner quartile range of 0.44 \mls\ to 1.32 \mls, and a median \mtmr\ of 0.15.  However, some low surface brightness dwarf irregular galaxies are know to have \hi\ mass-to-light ratios that are significantly higher than this quartile range.

One of the best known dwarf galaxies with a high \mlr\ is DDO~154, for which \citet{hof93} measured \mlr\ = 11\mls. DDO~154, a dwarf galaxy of type Sm, has a very extended \hi\ disk compared to its optical size, stretching to $\sim$6 times the Holmberg radius \citep{car98}. Another example is the Blue Compact Dwarf (BCD) galaxy NGC~2915, which has a moderately high \mlr\ of 2.6\mls\ \citep*{meu94,meu96}. The \hi\ disk of this object extends to over five times the Holmberg radius and has distinct spiral arm structure \citep[see][]{bur99}.  We do not consider \mlr\ ratios in tidal tails or bridges near galaxies, such as \hi~1225+01, a tidally disturbed irregular galaxy described in detail by \citet*{che95}.  Likewise, massive gas-rich low surface brightness galaxies such as Malin~1 \citep{bot87,imp89} are not comparable to low surface brightness dwarf systems with high \mlr\ ratios.

To date the largest study of galaxies with higher than average \hi\ mass-to-light ratios is that of gas-rich low surface brightness dwarf galaxies by \citet{vzee97} and \citet[and references therein]{vzee00,vzee01}.  This study included some galaxies with \mlr\ up to 7 \mls, as well as more "normal" higher surface brightness dwarf irregular galaxies (excluding BCDs).  \citet{vzee97} found that the star formation process in the low surface brightness and normal galaxies was inefficient, with the overall gas density well below the \citet{too64} instability threshold.  They measured some localized \hi\ peaks which approached the threshold and found that these were associated with local star forming regions.  Many of the galaxies in the later studies \citep{vzee00,vzee01} are isolated, with no neighbors found within $\sim$200--400 kpc.  \citet{vzee01} concluded that unlike Blue Compact Dwarfs, star formation within these galaxies was occurring at a slow, constant rate across the galaxy disk with few major concentrations of activity, and that gas depletion timescales for these galaxies were of the order of 20 Gyr.

While many galaxies with relatively high \mlr\ have been found, many more have gone unnoticed due in part to a bias towards optically selected samples and the limitations of past blind \hi\ surveys. The \hi\ Parkes All-Sky Survey (HIPASS) has provided the first ever blind survey of extragalactic neutral hydrogen in the local Universe ($v_{\rm sys} < 12~700$\kms) over the entire southern sky \citep{barn01}. One of its first products is the HIPASS {\it Bright Galaxy Catalog} \citep[hereafter BGC,][]{kor01,kor04}, which lists the 1000 \hi\ brightest sources in the southern hemisphere (selected by \hi\ peak flux density). This catalog gives us the opportunity to select interesting galaxies for detailed examination based on \hi\ and optical properties rather than entirely on optical characteristics.

\begin{figure} 
  \centering{\psfig{figure=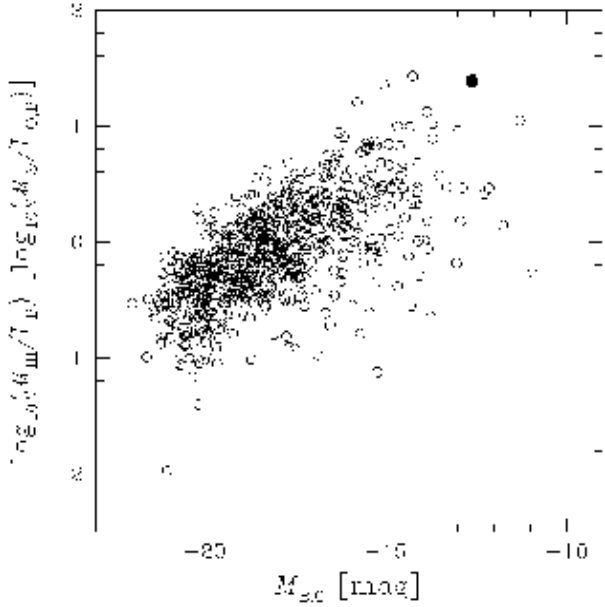,width=8cm}}
\caption{\hi\ mass-to-light ratio versus absolute photographic {\em B} 
   magnitude for 789 galaxies in the HIPASS Bright Galaxy Catalog 
   \citep{kor04} which have blue apparent magnitudes in LEDA. The 
   filled circle in the top right corner marks the position of \esoq.
\label{fig:mlmbt}}
\end{figure}

Photometric and structural parameters for the BGC's {\it optical} counterparts were obtained in 2002 from the Lyon-Meudon Extragalactic Database \citep[LEDA,][and references therein]{pat97} to study the statistical properties and various correlations of these galaxies.  First estimates of \hi\ mass-to-light ratios were obtained for 789 BGC galaxies which had mean photometric {\em B} band apparent magnitudes in LEDA.  Fig.~\ref{fig:mlmbt} shows the log(\mlr) distribution for these galaxies as a function of their absolute {\em B} magnitude:
\begin{equation} 
    M_{\rm B,0} = m_{\rm B} - A_{\rm B} - 5\log(D) - 25 ~~{\rm mag} ,
\end{equation}
where the galaxy distances, $D$ (Mpc), were calculated from the Local Group velocities given in the BGC.  Throughout this paper we adopt H$_0$ = 75\kms\,Mpc$^{-1}$.

Fig.~\ref{fig:mlmbt} shows a general trend for low luminosity galaxies in the BGC to have high \mlr.  The low luminosity and low \mlr\ region of Fig.~\ref{fig:mlmbt} is underpopulated as optically faint galaxies with little gas (dwarf elliptical galaxies) are not detected by HIPASS. We note that overall dwarf galaxies (those with the lowest luminosity) have a large spread in \hi\ mass-to-light ratios.  The lack of high \mlr\ bright galaxies, which should be easily detected, suggests that there is a physical limit to the \hi\ mass-to-light ratio for a galaxy of a certain luminosity (stellar mass).  Of the 148 galaxies in the BGC less luminous than $M_{B,0} = -16.5$ (hereafter referred to as dwarfs), 50 have \mlr\ $\ge$ 3\mls. Of these, seven galaxies have initial ratios greater than 10\mls\, and three have ratios in excess of 20\mls.  To understand the nature of these galaxies, their abundance of neutral hydrogen gas and low stellar content, we have started a multi-wavelength study of the galaxies with the highest \hi\ mass to light ratios in the southern sky, selected from the BGC. The filled circle in Fig.~\ref{fig:mlmbt} marks one of our sample galaxies, \esoq, with \mlr\ = $24 \pm 12$\mls, the second largest ratio in the BGC.

In this paper, Section~\ref{sec:eso215} looks at what was previously known about \esoq. Section~\ref{sec:obs} summarizes our observations.  Sections~\ref{sec:optp} and \ref{sec:radio} present the new results and properties of \esoq\ while Section~\ref{sec:dis} contains the discussion.  Section~\ref{sec:conc} gives our conclusions.

\section{The Galaxy \esoq} 
\label{sec:eso215}

\esoq\ was first catalogued on ESO(B) Atlas photographic plates \citep{lau82} and classified "G?", a {\em potential} galaxy.  It is necessary to retain the full original designation for ESO galaxies (including question marks) due to confusions that can sometimes arise between objects (such as with ESO174-G?001, ESO174-G001 and ESO174-GA001). \esoq\ is a dwarf irregular galaxy with a very low surface brightness. Because of its low Galactic latitude, $b$ = 10\fdg5, the Galactic extinction, \AB\ = $0.95 \pm 0.15$ mag \citep*{sch98}, is relatively high. Foreground star density in the field surrounding the galaxy is also high. No optical velocity is available to date. \citet{lau89} measured a total blue magnitude of $m_{\rm B} = 16.03 \pm 0.09$ mag from photographic plates.  No previous \hi\ observations are available other than those derived from HIPASS.  \citet*{huc01} published an \mlr\ for this galaxy (under the name KKs40) which was calculated using the \hi\ flux density from an early public data release of HIPASS and the \citet{lau89} photographic blue magnitude.  They found a high \mlr\ of $\sim12$\mls, comparable with DDO~154.  Unfortunately their \hi\ flux density is an underestimate as the public data release spectrum is off center and does not cover the entire galaxy.

Table~\ref{tab:prop} summarizes the properties of \esoq\ from previous studies, values derived from those quantities, and some of the results of this study. \citet{kor04} measured a systemic velocity of $v_{\rm sys} = 598 \pm 1$\kms, a velocity width of $w_{20} = 83 \pm 3$\kms, and an integrated \hi\ flux density of \FHI\ = $104.4 \pm 11.5$\jks.  We adopt a distance to \esoq\ of $D$ = 4.2 Mpc from the Local Group velocity of $312 \pm 1$\kms\ using $D = v_{\rm LG}/H_0$ (see also Section~\ref{sec:dis-uncert}).

The proximity and isolation of \esoq\ make it an excellent candidate for a study of the nature of high \hi\ mass-to-light ratio galaxies, and of galaxy evolution in general. Using deep optical {\em BVRI} band images and \hi\ data of \esoq\ we will examine the stellar and \hi\ distribution, stellar population, isolation and the \hi\ kinematics to search for the physical reasons why \esoq\ could maintain such a high fraction of \hi\ gas, while other galaxies processed most of their gas into stars.

\section{Observations and Data Reduction} 
\label{sec:obs}
The galaxy \esoq\ was observed in two different wavelength regimes. Optical photometry was obtained with the Australian National University (ANU) 2.3-meter Telescope at the Siding Springs Observatory. \hi\ line and radio continuum data were obtained with the Australia Telescope Compact Array (ATCA).

\subsection{Optical Photometry} 

Optical {\em BVRI} band (Cousins filters) images were taken as a series of 300\,s exposures in April/June 2002 and May 2003 using the Nasmyth Imager (SITe $1124 \times 1024$ thinned CCD). The imager has a circular field of view with a diameter of 6\farcm62 and a pixel size of 0\farcs59. The total exposure time in the four bands is 50, 40, 30, and 30 minutes, respectively. The typical seeing in the final images for the different bands was $\sim$1\farcs9, 1\farcs9, 1\farcs8, and 2\farcs0, respectively, and most observations were taken at low airmass. Twilight sky flat fields in all bands and bias images were obtained at the same time. Several \citet{lan92} standard stars were taken together with shallow 120\,s {\em BVRI} images of the galaxy field on a photometric night to perform the photometric calibration of the deeper images.

Data reduction and analysis was carried out with IRAF using standard procedures.  After overscan subtraction, bias subtraction, and flatfielding, individual 300s images were registered to each other and the sky level was subtracted. The images for each band were then combined into a single image (to increase signal-to-noise, remove cosmic rays, etc.) and the photometric calibration applied.

Foreground stars were removed by replacing them with the surrounding sky so that only the galaxy remained.  Special care was taken to restore the light distribution under stars superimposed onto the galaxy, using the mirror image from across the galaxies center.  Fig.~\ref{fig:optimage} shows the resulting {\em BVRI} images before and after star subtraction. We measured the luminosity-weighted center of the galaxy (averaged between the four bands) to be at $\alpha,\delta$(J2000) = $10^{\rm h}\,57^{\rm m}$\,29\fs9, $-48$\degr\,10\arcmin\,43\arcsec ($\pm$2\arcsec), consistent with \citet[see Table~\ref{tab:prop}]{lau82}.

\subsection{Radio Observations} 

\hi\ line and 20-cm radio continuum observations of \esoq\ were obtained between June 2002 and March 2003 with the EW352, 750A and 6A arrays for $\sim$12 h each. The band was centered at a frequency of 1417 MHz with a bandwidth of 8 MHz, divided into 512 channels. The channel width is 3.3\kms, and the velocity resolution is $\sim$4\kms. The primary and secondary calibrators were PKS 1934--638 (14.88 Jy) and PKS 1215--457 (4.8 Jy), respectively.

Data reduction and analysis was performed with the ${\cal MIRIAD}$, ${\cal AIPS}$, and KARMA packages using standard procedures. Channels with Galactic emission were discarded. After continuum subtraction, the \hi\ data were Fourier-transformed using `natural' weighting and a channel width of 4\kms. The data were cleaned and restored with a synthesized beam of $38\arcsec \times 35\arcsec$. We measure an r.m.s. noise of 1.6 mJy per channel, close to the theoretical value. Primary beam correction was applied.  \hi\ moment maps were obtained using cutoffs of 4 mJy beam$^{-1}$ (\hi\ distribution) and 6 mJy beam$^{-1}$ (mean velocity field and dispersion).

Low resolution 20-cm radio continuum maps were produced using multi-frequency synthesis and `natural' weighting resulting in a synthesized beam of $18\arcsec \times 17\arcsec$ and an r.m.s. of $\sim$0.13 mJy. Primary beam correction was applied. We do not find any radio continuum emission associated with \esoq. We estimate the 20-cm flux density to be less than 1 mJy over the optical extent of the galaxy. To obtain the star formation rate we used the relation by \citet{con92} and \citet{haa00}:
\begin{equation}  
SFR = 0.14~D^{2}~S_{{\rm 20cm}} ~~{\cal M}_{\sun}\,{\rm yr}^{-1},
\label{eqn:sfr}
\end{equation} 
where $D$ is the distance in Mpc and $S_{{\rm 20cm}}$ is the 20-cm radio continuum flux density in Jy. We calculate an upper limit to the star formation rate for \esoq\ is $\sim 2.5 \times 10^{-3}$\Msun\,yr$^{-1}$.

\section{Optical Properties} 
\label{sec:optp}

Fig.~\ref{fig:optimage} shows the deep {\em BVRI} band images of \esoq\ before and after subtraction of the foreground stars.  A growth curve of the galaxy was measured for the star-subtracted images from the center in 2 pixel ($\sim$1\farcs2) circular aperture rings to obtain the total intensity.  We are confident that the sky background subtraction was successful as the growth curves for all bands flattened out beyond the galaxy at around the same radius.  From these the empirical apparent magnitude, effective radius and effective surface brightness for each band were calculated.  The main error for the apparent magnitude is the uncertainty in the background sky level.  The derived {\em B} band  magnitude of \mB\ = $16.13 \pm 0.07$ mag (within 80\arcsec\ of the center) is consistent with \citet[see Table~\ref{tab:prop}]{lau89}.  The {\em R} band magnitude of \mR\ = $14.38 \pm 0.11$ mag is however 0.64 mag brighter than that from \citet{lau89}.  From \mB\ and our adopted distance, \esoq\ has an absolute magnitude of \MB\ = $-12.9 \pm 0.2$ mag.  The $B$ luminosity is given by:
\begin{equation}  
L_{\rm B} = D^{2}~10^{10 - 0.4(m_{\rm B}-A_{\rm B}-M_{{\rm B}, \sun})} ~~L_{\sun,{\rm B}},
\label{eqn:lum}
\end{equation} 
where the absolute Solar {\em B} magnitude, $M_{{\rm B}, \sun}$, is taken as 5.48 mag \citep*{bes98}.  We derive a {\em B} band luminosity of \LB\ = $(2.3 \pm 0.4) \times 10^7$\Lsun, a factor of $\sim$80 below the median {\em B} luminosity of $1.9 \times 10^{9}$\Lsun\ for Sm/Im galaxies in \citet{rob94}.

We then obtain surface brightness profiles from all four bands, shown in Fig.~\ref{fig:optprofile}.  A S\'ersic profile \citep{ser68}, which is a generalized version of an exponential profile, was fit to all the surface brightness profiles.  The S\'ersic law, in terms of surface brightness $\mu(r)$, is defined by:
\begin{equation}  
\mu(r) = \mu_{0} + 1.086(r/r_{0})^n ~~{\rm mag~arcsec}^{-2},
\label{eqn:sersic}
\end{equation} 
where $r$ is the radius from the center in arcsec, $\mu_{0}$ is the central surface brightness in mag arcsec$^{-2}$, $r_{0}$ is the scale radius in arcsec, and $n$ is the shape parameter (where $n = 1$ gives an exponential profile).  These profiles are not normally fit to dwarf irregular galaxies as they commonly have clumpy \hii\ regions on top of an exponential disk, but as \esoq\ is much smoother than most late-type galaxies it is well described by a S\'ersic model.  The S\'ersic fits were used to extrapolate beyond the detection limit to get the total apparent magnitude.  The magnitudes are consistent with the empirical magnitudes derived from the total counts above the sky level, and the empirical values are adopted for the remainder of the paper.  The radial extent of \esoq\ at an extinction corrected {\em B} band surface brightness of 26.6 mag arcsec$^{-2}$ (the Holmberg radius) is $57\farcs6 \pm 0\farcs6$.

To the north in the optical images ($\sim 2.5\arcmin$ from the center) there is a diffuse patch of stars which may be related to star formation in the galaxy, which is on the eastern edge of the strongest peak in the \hi\ distribution.  Unfortunately there is also a strong concentration of foreground stars on the eastern side of this patch, including the brightest star in the field, which makes the star subtraction here more difficult and increased the uncertainty for any luminosity determination.

Table~\ref{tab:mag} gives a summary of the photometric results obtained from these profiles.  No Galactic extinction correction was applied to the apparent magnitude and surface brightness.  Column (1) gives the Cousins broad band filters used.  Column (2) gives the empirically derived total apparent magnitude.  Column (3) gives the total apparent magnitude as extrapolated from the S\'ersic profile fit.  Column (4) gives the central surface brightness. Column (5) gives the effective surface brightness, the average surface brightness out to the half light radius, and Column (6) gives the half light (effective) radius.  Column (7) gives the radii out to $\mu = 26.6$ mag arcsec$^{-2}$ (which is the Holmberg radius in the {\em B} band, extinction corrected).  Column (8) gives the Galactic extinction correction from \citet{sch98}.

We obtained six colors from the photometry, which are shown at the end of  Table~\ref{tab:typical} (which we will discuss in Section~\ref{sec:dis-mlratio}).  From these colors it seems that \esoq\ is very red for a late-type galaxy, suggesting that the stellar population is older than is typically seen in dwarf irregulars (see discussion in Section~\ref{sec:dis-starpop}).  Fig.~\ref{fig:colourprofile} shows the color profiles for each of these six color combinations.

\section{Radio Properties} 
\label{sec:radio}

\subsection{\hi\ Structure} 
\label{sec:structure}

Fig.~\ref{fig:hichannel} shows the \hi\ channel maps of \esoq.  \hi\ line emission was detected in the velocity range 547 to 647\kms.  The emission has a resemblance to a slowly rotating gas disk, except that the pattern is not as sharply defined as might be seen in a larger disk \citep[see example in ][]{gio88}.  Also, the structure throughout the channels is not evenly distributed with several concentrations and voids.  In particular, to the north of the galaxy center the gas has two strong peaks close to the systemic velocity, and to the south east of the center a depression is visible around 609\kms.  We find no companion galaxies within the ATCA primary beam (33\arcmin) and the velocity range of 200 -- 1300\kms.

Fig.~\ref{fig:hispectra} shows the \hi\ spectra of \esoq\ as obtained from the HIPASS BGC and the ATCA. There is a resemblance to a double horn profile, characteristic of a rotating disk, although the central dip is not pronounced (which may be partially due to the low inclination of the source).  Notably the ATCA spectrum is inconsistent with the single dish result, and measures a higher flux density.  The HIPASS BGC flux density may be underestimated due to the large extent and high peak flux density of the source \citep[see][]{barn01}.

Fig.~\ref{fig:mom} shows the integrated \hi\ distribution, the mean \hi\ velocity field and the \hi\ velocity dispersion of \esoq.  From the \hi\ distribution we measure the dimensions of the \hi\ disk as $730\arcsec \times 640\arcsec$ ($\pm$40\arcsec) at a position angle of $PA \sim 120$\degr.  Fig.~\ref{fig:hirad} shows the azimuthally averaged radial \hi\ density profile of \esoq.  The profile was measured with rings 10\arcsec\ wide assuming an inclination of 36\degr\ and PA $=$ 119\degr\ (Section~\ref{sec:velo}).  Here 1 Jy beam$^{-1}$\kms\ corresponds to a column density of $8.32 \times 10^{20}$ atoms\,cm$^{-2}$.  \hi\ can be traced out to a column density of $\sim 5.0 \times 10^{19}$ atoms\,cm$^{-2}$ at a radius of $370\arcsec \pm 20\arcsec$.  We obtain an \hi\ flux density of \FHI\ = $122 \pm 4$\jks.  This is higher than the \FHI\ = $104.4 \pm 11.5$\jks\ obtained in the BGC \citep{kor04}.  Using the adopted distance of 4.2 Mpc we derive an \hi\ mass of \MHI\ = $(5.1 \pm 0.2) \times 10^{8}$\Msun.  This is less than half the median \hi\ mass of $1.27 \times 10^{9}$\Msun for Sm/Im galaxies in \citet{rob94}, but is still well within the range of \hi\ masses seen in this sample (see Table~\ref{tab:typical}).

\subsection{\hi\ Gas Dynamics} 
\label{sec:velo}

Overall the mean \hi\ velocity field (Fig.~\ref{fig:mom}c) is remarkably regular, especially for an late-type galaxy where the velocity field is usually extensively disturbed \citep*[see examples in][]{cot00,sti02b}.  There are few asymmetries between the two sides of the galaxy.  The last contour on the approaching side is closed while the receding side contours are still open, which suggests the rotation curve on the approaching side may turn down.  There is also some asymmetry across the major axis of the galaxy, most clearly seen in the channel maps (Fig.~\ref{fig:hichannel}) around $v_{hel} \sim 609$\kms\ where the southern portion of the galaxy curves around more than the northern portion.  However, on the whole this is one of the most symmetric and undisturbed velocity fields of a low-mass, late-type galaxy.

Through most of the galaxy the velocity dispersion (see Fig.~\ref{fig:mom}d) is between about 4 to 10\kms.  In the central region around the stellar component of the galaxy, and extending to the north towards the strongest \hi\ concentration, the velocity dispersion is largest.  This is reflected in Fig.~\ref{fig:veldisp}, which shows the azimuthally averaged radial \hi\ velocity dispersion profile of \esoq.  The velocity dispersion is highest in the center of the galaxy where the profile is peaked.  In the outer galaxy the dispersion is lower with a slightly declining profile.  Fig.~\ref{fig:pvd} shows an \hi\ position velocity diagram along the major axis of \esoq, emphasizing the symmetry between the approaching and receding sides of the galaxy.  On the receding side there are two gaps in the gas distribution accompanied by a large spread in velocity.  These features could be expanding bubbles which have been recognized in other face-on galaxies \citep*[such as the large \hi\ `superbubble' in M101,][]{kam91}.  They correspond to the low density regions in Fig.~\ref{fig:mom}b to the south east.  There is no evidence of a central \hi\ bar in Fig.~\ref{fig:pvd}, however a weak bar on the scale of the optical component of the galaxy would be difficult to pick up from this diagram with our current resolution.

We obtained a rotation curve from the mean \hi\ velocity field of \esoq\ using the tilted ring algorithm in {\cal AIPS} \citep[][ {\sc rocur}]{beg89}. Using all data within a radius of 6\arcmin, we derive the dynamical center of the galaxy at $\alpha,\delta$(J2000) = $10^{\rm h}\,57^{\rm m}$\,30\fs0, --48\degr\,10\arcmin\,47\arcsec\ ($\pm$10\arcsec), and its systemic heliocentric velocity, $v_{\rm sys} = 597 \pm 1$\kms.  This is consistent with \citet{kor04}, and the position agrees with that derived from the center of the stellar component. The center and systemic velocity were fixed for the fitting of the remaining parameters for which we used 12 rings of 35\arcsec\ width. 

We found the position angle of \esoq\ to be constant with radius and over both receding and approaching sides at $PA = 119\degr \pm 2\degr$, and so was fixed for the remainder of the fitting.  The inclination angle $i$ varied between 34\degr\ and 44\degr\ beyond 180\arcsec, but remained mostly around 36\degr\ and was fixed at this angle to fit the rotation curve.  The resulting rotation curve is shown in Fig.~\ref{fig:rotc}.

Individual analysis of the approaching and receding sides of the galaxy revealed some differences.  The inclination for the best fit to the outer part of the galaxy on the receding side was $i = 40\degr \pm 6\degr$ (beyond 150\arcsec), while for the approaching side it was $i = 31\degr \pm 4\degr$.  Fig.~\ref{fig:rotc} includes the rotation curves for both sides fit with the same inclination angle as the overall fit, 36\degr.  There is a slight turn down on the approaching side, while the receding side continues rising slightly.  Overall the three rotation curves are reasonably flat beyond 150\arcsec.

With $i = 36\degr \pm 10\degr$ we find a rotational velocity of \vmax\ = $51 \pm 8$\kms\ out to a radius of $370\arcsec \pm 20\arcsec$ ($\sim$7.5 kpc). We derive a total dynamical mass of \Mtot\ = $(4.5 \pm 1.6) \times 10^9$\Msun.  This gives an \hi\ to dynamical mass ratio of \mtmr\ = $0.11 \pm 0.04$, at the low end of what \citet{rob94} found for Sm/Im galaxies (median value 0.15).

\section{Discussion} 
\label{sec:dis}

\subsection{Comparing Optical and \hi\ Properties}
\label{sec:dis-opthi}

Using Eqn.~\ref{eqn:mlr}, we can now recalculate \mlr\ using the ATCA integrated \hi\ flux density of $122 \pm 4$\jks, the {\em B} band apparent magnitude of $16.13 \pm 0.07$ mag, and the \citet{sch98} Galactic extinction of $0.95 \pm 0.15$ mag.  We derive an \hi\ mass-to-light ratio for \esoq\ of $22 \pm 4$\mls.  A large portion of the uncertainty is due to the uncertainty in the Galactic extinction.  This is in good agreement with the preliminary ratio from the BGC of $24 \pm 12$\mls.  To our knowledge \esoq\ has one of the highest (if not the highest) \mlr\ to be confirmed by accurate measurement to date for any galaxy system, being approximately double the \mlr\ of DDO~154 \citep{hof93}.  Using \Mtot\ = $(4.5 \pm 1.6) \times 10^9$\Msun\ and \LB\ = $(2.3 \pm 0.4) \times 10^7$\Lsun, the dynamical mass-to-light ratio is \tmlr\ = $200 \pm 110$ \mls, making \esoq\ an extremely dark matter dominated galaxy.  If we assume a stellar mass-to-light ratio of $\sim$1\mls, we estimate a baryonic mass to total mass ratio of \bmtmr\ = 0.12.

Fig.~\ref{fig:mom}a shows that the \hi\ envelope extends out much further than the main stellar component of the galaxy.  For \esoq\ the \hi\ gas can be traced out to $370\arcsec \pm 20\arcsec$ while the Holmberg radius is only $57\farcs6 \pm 0\farcs6$ (at the extinction corrected surface brightness of 26.6 mag arcsec$^{-2}$).  Therefore the gas extends to $6.4 \pm 0.4$ times the Holmberg radius, comparable to the extent of DDO~154 \citep{car98}.

The brightest part of the central stellar region (see Fig.~\ref{fig:optimage}) is slightly elongated in a direction aligned much closer to the rotational minor axis ($\sim$40\degr).  As the galaxy is seen at a low inclination this feature is potentially a weak central bar.  The optical morphology of \esoq\ is otherwise rather smooth and featureless for a dwarf irregular, although its slightly mottled appearance in color images, the slowly rotating \hi\ disk, and possible \hii\ regions (see Section~\ref{sec:dis-starpop}) imply it is a late-type dwarf galaxy.  The stellar distribution is located in the center of the \hi\ disk at the position of a shallow depression in the \hi\ distribution.

\subsection{The Importance of Accurate Multi-wavelength Measurements}
\label{sec:dis-orion}

To demonstrate the great uncertainties that can arise in determining \hi\ mass-to-light ratios we will briefly discuss the Orion dwarf galaxy (\hi~0542+05).  In their initial study of \hi\ envelopes around low luminosity galaxies, \citet*{vzee95} included 33 galaxies mostly taken from the {\em Uppsala General Catalog} which they believed to have high \mlr\ ($>$ 5 \mls).  One of these galaxies is the Orion dwarf, for which they derived an extreme \mlr\ of 84.1\mls\ using their measured Arecibo flux density of \FHI\ = 74.16\jks\ and an extinction corrected {\em B} magnitude of 17.1 mag.  They did not discuss this galaxy further.  The optical magnitude for this galaxy, and for many others objects in their sample, had been obtained by eye.  Unfortunately many were underestimates.  \citet{vzee97} states that "...the catalogued magnitudes of these systems were a severe underestimate of their true luminosity (by $\sim 1.5^{m}$); however, the revised values of ${\cal M}_{H}/L_{B}$ are still more than a factor of 2 higher than typical for dwarf galaxies".  In later investigations \citep{vzee96,vzee97,vzee00,vzee01} no galaxies with \mlr\ $>$ 7 \mls\ were found.  In addition to this the Galactic extinction correction used for the Orion dwarf galaxy \citep[$\sim$0.9 mag,][although it is not clearly stated]{bur78} is possibly also a severe underestimate as recent measurements give \AB\ = $3.17 \pm 0.51$ mag \citep{sch98}, and \citet{bur82} give 2.84 mag.  They may have mistaken \AB\ for $E{B - V}$.  \citet{kar96} performed the first CCD photometry on this galaxy, getting a new \mB\ of 15.7 mag, and using an extinction of 2.7 mag derived \mlr\ = 2.6\mls.  \citet*{bar01} found a similar {\em B} magnitude of 15.41 mag.  Combined with the rise in the Galactic extinction, this galaxy appears approximately 5 magnitudes brighter than in \citet{vzee95}, which means \mlr\ is overestimated by a factor of around 100.  Using the \citet{bar01} {\em B} magnitude, the \citet{sch98} Galactic extinction, and the \hi\ flux density from HIPASS (\FHI\ = 73 $\pm$ 4\jks), a more accurate \hi\ mass-to-light ratio for the Orion dwarf (HIPASS J0545+05) is $0.9 \pm 0.5$\mls.

Problems can also arise with the \hi\ observations.  \citet*{one00} find four galaxies in the range 10\mls\ $<$ \mlr\ $<$ 50\mls\ as part of an Arecibo follow-up of a group of low surface brightness galaxies, as well as six galaxies which deviated from the Tully-Fisher relationship.  These were the first extragalactic observations taken after an upgrade at Arecibo.  Later \citet{chu02} found that many of the galaxies observed by \citet{one00} had been contaminated by other neighboring objects within the sidelobes \citep[see][]{hei01}.

We conclude that to obtain accurate \hi\ mass-to-light ratios it is vitally important that the \hi\ flux density, the {\em B} band apparent magnitude and the Galactic extinction are well determined; error bars on all are essential and should always be quoted.  Galaxies located near the Galactic Plane, where the extinction is high and uncertainty increases, are particularly difficult and should be treated with extra care.

\subsection{Other Uncertainties to Consider}
\label{sec:dis-uncert}

How well do we know the distance to \esoq?  While we have adopted the Hubble flow distance to \esoq\ of 4.2 Mpc using the Local Group velocity in this paper, the relative uncertainty in such a distance can be large for nearby galaxies.  This is mostly due to peculiar motions and the difficulty in measuring the Hubble constant in the Local Universe.  Alternatively, we can calculate a distance from the Baryonic Tully-Fisher relationship \citep{mcg00} by first estimating a baryonic mass for \esoq\ from the relationship given the galaxy's rotation velocity of $51 \pm 8$\kms.  We estimate \Mbary\ = $2.1 \times 10^{8}$\Msun.  Given the \hi\ flux density and the {\em B} band apparent magnitude then \esoq\ would have to be at a distance of 2.6 Mpc to have this baryonic mass, almost half the Hubble flow distance.  At the Hubble flow distance \esoq\ would lie above the Baryonic Tully-Fisher relationship.  Table~\ref{tab:distdep} gives examples of the effect of this uncertainty on the main distance dependent properties in this paper.  It shows the derived quantities at both distances, with the quoted uncertainty accounting for all other sources of error.  Uncertainty in the estimated distance can have a large effect on properties such as {\em B} band luminosity and \hi\ mass, both of which change by a factor of $\sim$2.5 between the two estimates.  The Baryonic Tully-Fisher distance was not adopted in this paper as there is currently large uncertainty at the low mass end of the relation.

Galactic dust extinction can strongly affect the final \mlr\ result, as we have seen in Section~\ref{sec:dis-orion}.  The relatively high Galactic extinction in the direction of \esoq\ and the high number of foreground stars may mean that we could ultimately miss some very faint components of the galaxy, and so underestimate the total apparent magnitude.  There is also the question of whether the value of Galactic extinction we are using is completely wrong.  The \citet{sch98} Galactic extinction is measured over large scales, and there may be smaller scale variation.  The redder color of the galaxy may suggest stronger extinction than we are accounting for.  To test this we looked at the extinction correction for galaxies in the surrounding field to see if there was a large extinction gradient in the area.  We found that out to a radius of 1\degr\ the variation was between 0.8 and 1.1 mag, within the uncertainties of the extinction at \esoq's position, and so we are confident that \AB\ is not too far from the \citet{sch98} value.  However we note that close to the Galactic Plane the dust distribution can be patchy, and it is entirely possible that the extinction could be much higher at the position of \esoq\ than current measurements indicate.

Extinction from within \esoq\ itself must also be considered as that may reduce the total light we receive and redden the galaxy.  It is not possible currently to estimate what the dust extinction could be in this galaxy as internal extinction is poorly understood in dwarf irregular galaxies, though it is often assumed to be low.  The inclination of the galaxy is quite low as seen from the rotation curve analysis, and assuming that most of the galaxy's dust is in the plane of the \hi, we find it unlikely that internal dust extinction has a large effect on the luminosity of \esoq.

\subsection{Stellar Population and Star Formation}
\label{sec:dis-starpop}

The photometric colors (see end of Table~\ref{tab:typical}) are all fairly red for a dwarf irregular galaxy, suggesting that \esoq\ contains a mainly old stellar population, which is uncommon in late-type galaxies.  \citet*{par02} find the mean color for late-type galaxies in their sample is $\langle B - R \rangle _{Late} = 0.90$, while for the small number of early-types in their sample it is $\langle B - R \rangle _{Early} = 1.21$.  Their reddest late-type galaxy is ESO006--G001 with $B - R = 1.24$, a galaxy they claim has an intermediate morphology (both late- and early-type).  \esoq's value of $(B - R)_{0} = 1.39 \pm 0.22$ is closer to the colors of the dwarf ellipticals and the more extreme late type dwarfs in the sample.  The median $B - V$ color for 210 Sm/Im galaxies in the Local Supercluster from \citet{rob94} is 0.42.  Again \esoq\ has a much redder color, $(B - V)_{0} = 1.02 \pm 0.21$.  It should be noted that the uncertainties in these colors are quite high due to the position of the galaxy close to the Galactic Plane.  Unfortunately these higher uncertainties due to Galactic extinction prevent us from performing a detailed analysis of the age and metalicity of the underlying stellar population.

Several of the color profiles (Fig.~\ref{fig:colourprofile}) show slight gradients between the inner and outer regions of the galaxy.  The $(B - I)_{0}$, $(V - I)_{0}$, and $(R - I)_{0}$ profiles all have shallow but distinct positive gradients going out from the galaxy center, meaning that the outer region of the galaxy is redder than the center.  $(B - V)_{0}$, $(B - R)_{0}$, and $(V - R)_{0}$ had mostly flat trends, with the first two turning slightly to the blue at the outer edges.  \citet{par02} also find that dwarf galaxies often have a positive color gradient, with younger/bluer stars in the inner regions and the older/redder population in the outer regions.  This is the opposite to larger spiral systems where spiral density waves drive star formation in the outer disk and the central region is dominated by an older red bulge.  Of the late-type galaxies in the \citet{par02} sample only eight show a positive gradient in $B - R$, while a further eight showed a flat profile and a few showed negative profiles.  In this color \esoq\ also shows a mostly flat trend.  It appears that there may be a slightly higher proportion of younger stars in the center of \esoq, but as we don't see it in all colors this trend is unlikely to be strong.  \citet{vzee01} find that color gradients in dwarf irregular galaxies in a quiescent phase (non-starbursting) are generally not as pronounced as those of starburst galaxies (such as BCDs, where there is strong star formation confined to a small central region).

During the star subtraction of the optical images (Fig.~\ref{fig:optimage}) several bright slightly extended objects close to the galaxy which we believe are \hii\ regions were not removed.  The two most prominent can be seen best in the {\em B} image, on the outer north-eastern edge of the galaxy and close in on the north-western side.  These regions appear to be rather blue compared to the rest of the galaxy.  However they may be background galaxies or foreground features rather than \hii\ regions associated with \esoq.  To try and confirm that these are indeed star forming regions we examined the SuperCOSMOS H-alpha Survey (SHS) image of the field around \esoq\ \citep[and references therein]{ham01}.  The image can be seen in Fig.~\ref{fig:20cmha} with the contours from our 20-cm radio continuum observation of the field overlaid.  The two regions mentioned show up distinctly as diffuse objects, while the main galaxy is barely visible at all.  Two further diffuse regions are visible well beyond the optical galaxy, and almost to the edge of the \hi\ disk on the south-eastern side.  However, none of these diffuse regions in the H$\alpha$ image correspond to any of the 20-cm continuum sources in the field.  All of these objects could be \hii\ regions in \esoq, but this needs to be confirmed with follow up spectra and H$\alpha$ imaging.

The upper limit for the star formation rate as derived from the 20-cm radio continuum flux density measurement within the optical galaxy region, $\sim 2.5 \times 10^{-3}$\Msun\,yr$^{-1}$, is on the lower side even when compared to other quiescent dwarf irregular galaxies such as the sample of \citet{vzee01}.  As is described in that paper, deriving a past star formation rate to compare to the present rate can be problematic for dwarf irregular galaxies.  A rough estimate of the average past star formation rate for \esoq\ can be derived by dividing the total mass of stars formed by the length of time for which the galaxy has been forming stars, which can be expressed as \citep{tin80}:
\begin{equation}  
\langle SFR \rangle_{past} = \frac{\Gamma_{\rm B} L_{\rm B}}{T_{sf} (1 - R)} ~~{\cal M}_{\sun}\,{\rm yr}^{-1},
\label{eqn:sfrpast}
\end{equation} 
where $\Gamma_{\rm B}$ is the {\em B}-band stellar mass-to-light ratio in \mls, $T_{sf}$ is the time for which the galaxy has been forming stars in years, and $R$ is the recycling fraction (essentially the proportion of the stars formed within the galaxy that are no longer visible).

$\Gamma_{\rm B}$ is poorly understood for dwarf irregular galaxies \citep[see][and references therein]{vzee01}, and as a conservative estimate we used $\Gamma_{\rm B} \sim 1$ \mls\ for \esoq, although the actual value is probably higher given the seemingly older stellar population of the galaxy.  For the other parameters in Eqn.~\ref{eqn:sfrpast} we use the assumptions made in \citet{vzee01} of $T_{sf} = 13$ Gyr and $R = 0.33$.  Using \LB\ = $(2.3 \pm 0.5) \times 10^7$\Lsun\ we derive $\langle SFR \rangle_{past} \sim 2.7 \times 10^{-3}$\Msun\,yr$^{-1}$.  Although this is uncertain, it is close to our upper limit for current star formation rate, suggesting that either star formation has been more or less constant throughout \esoq's history, or it may have been slightly higher in the past.  It certainly implies that there have not been any major starburst events in \esoq\ during its history, consistent with the galaxy being spatially isolated and possibly never having interacted with another galaxy.  The minimum gas depletion timescale for \esoq\ given the current star formation rate upper limit and using the same recycling fraction as above is $\sim 300$ Gyr.

\subsection{The Star Formation Threshold}
\label{sec:dis-toomre}

\citet{ken89} and \citet{mar01} (hereafter K89 and MK01, respectively)  have found that gravitational instability may determine the star formation law for galactic disks.  K89 found that a modified \citet{too64} Q criterion, a simple thin disk gravitational stability condition, could satisfactorily describe the star formation threshold gas surface densities in active star-forming galaxies.  From this criterion they derived a critical gas surface density $\mu_{crit}(r)$, given by (MK01):
\begin{equation}  
\mu_{crit}(r) = \alpha_{Q} \frac{\sigma(r) \kappa(r)}{\pi G},
\label{eqn:mucrit}
\end{equation} 
where $\sigma(r)$ is the gas velocity dispersion, $\kappa(r)$ is the epicyclic frequency, and $\alpha_{Q}$ is a constant close to unity which is included to account for a more realistic disk (i.e. thicker disk, embedded stellar population).  K89 derive $\alpha_{Q} = 0.63$\footnote{Adjusted as in MK01}, while for the MK01 sample (which was larger and had more detailed input parameters) they found $\alpha_{Q} = 0.69$.  If $\mu_{gas}(r)$, the surface density of the gas in the disk, exceeds $\mu_{crit}(r)$ then the disk will be unstable to axisymmetric disturbances and large scale star formation can occur.  MK01 also found that the simple \citeauthor{too64} criterion breaks down in some circumstances, such as low shear rates and stellar bars, and that there may be a weak trend for late-type galaxies to have higher $\alpha_{Q}$.

We have examined the star-formation threshold throughout \esoq\ as a function of radius to determine if a subcritical gas density is preventing the galaxy from large scale star formation.  The gas velocity dispersion $\sigma(r)$ comes from our azimuthally averaged \hi\ velocity dispersion radial profile.  The epicyclic frequency $\kappa(r)$ can be determined from our \hi\ rotation curve using (MK01):
\begin{equation}  
\kappa^{2}(r) = 2\left( \frac{V^2}{R^2} + \frac{V}{R} \frac{dV}{dR} \right) ~~{\rm s}^{-2},
\label{eqn:epkap}
\end{equation} 
where $V$ is the rotation velocity in \kms, and $R$ is the radius in km.  The derivative was determined with a forward difference method.  We have assumed $\alpha_{Q} = 1$.  Only the azimuthally averaged \hi\ surface density ($\mu_{\rm HI}(r)$) was available for the gas surface density.  We have made no estimate of the molecular hydrogen content (as was done in K89) for \esoq\ at this stage.

From these data we have estimated the ratio $\mu_{\rm HI}(r)/\mu_{crit}(r)$ as a function of radius, which can be seen in Fig.~\ref{fig:toomreq}.  Over all radii the ratio is less than the MK01 value of $\alpha_{Q} = 0.69$, strongly suggesting that the \hi\ gas surface density of \esoq\ is not high enough to induce large scale star formation within the disk. The profile is similar to the galaxies in both K89 and MK01 with low levels of star formation.  However, the majority of those galaxies are early-type spiral galaxies (Sa to Sb), not dwarf galaxies.  The exclusion of molecular gas from our estimate may have caused a significant underestimate of $\mu_{gas}(r)/\mu_{crit}(r)$, which may lead to part of the profile rising above the threshold.  The estimate which K89 used ($\mu_{gas}(r) = 1.41 \mu_{\rm HI}(r)$) may not be appropriate for \esoq\ or other dwarf irregular galaxies, so the only way to resolve this issue would be to obtain CO observations to estimate the H$_2$ content (which may be problematic for \esoq\ if the metalicity and excitation are low).  It is important to note also that, as MK01 mention, using the azimuthally averaged \hi\ surface density may not be appropriate for galaxies like \esoq\ where the variation in density at a particular radius can be large.  Several regions of the galaxy have \hi\ densities much higher than the azimuthal average at their location (such as the region to the north of the galaxy center) and local densities may approach or exceed the threshold density.

\subsection{Possible Causes of a High \mlr}
\label{sec:dis-mlratio}

Table~\ref{tab:typical} gives a summary of the major properties of \esoq, and how they compare to literature values for "typical" late-type dwarf galaxies.  Column (1) gives the property.  Column (2) gives the value we have derived for \esoq.  Column (3) gives the typical range for this property in late-type dwarf galaxies as found in various literature sources.  Column (4) gives the reference for the column (3) value, and Column (5) gives the galaxy types within the sample from that reference.  The results from \citet{rob94} are given as the inter-quartile range for Sm and Im galaxies, while other results cover the full range of the sample given in the reference.  When compared to the \citet{rob94} sample, the total mass of \esoq\ is at the low end of the quartile range, and we can use this as an anchor point for comparing the other parameters.  We see that the \hi\ mass is also at the low end of the quartile range (along with other parameters not dependent on the luminosity), and in general is consistent with the literature values for late-type galaxies.  However, along with \esoq's \mlr, parameters such as luminosity (and therefore absolute magnitude) and total mass-to-light ratio can differ from the typical dwarfs by an order of magnitude or more.  Galaxies with such low luminosities/absolute magnitudes are still in other dwarf irregular samples \citep[see][]{lee03,sti02a} but are generally among the outliers.  This suggests that \esoq\ has a normal \hi\ content for its type, but is underluminous.

\esoq\ shows no signs that it has been perturbed by a neighbor.  No nearby galaxies were found within the \hi\ data cube.  We searched the field around the galaxy in LEDA out to approximately 13\fdg6 (1 Mpc at our assumed distance of 4.2 Mpc) within the velocity range 200\kms\ $< v_{sys} <$ 1300\kms.  We found nine galaxies, all with \MB\ $<$ -17.5 mag, all but one being a late-type spiral or irregular. Only one galaxy, ESO264--G035, had a systemic velocity close to \esoq's of $v_{sys} = 770$\kms, which is still much higher than \esoq's velocity.  \citet{kar02} include \esoq\ (KKs 40) in their list of galaxies around the Centaurus A group (26\fdg16 away).  While this is the nearest group, assuming that they are the same distance from us \citep*[the Cen A group is $\sim 4$ Mpc away,][]{jer00} then \esoq\ is well outside the group at a projected distance of 1.9 Mpc.  From our current knowledge of the field around it \esoq\ is a very isolated system.

If \esoq\ were still in its first episode of collapse that may explain why it has been able to hold onto its \hi.  For the rotational velocity at the outer edge of the \hi\ disk (see Section~\ref{sec:velo}) the dynamical timescale of \esoq\ is $9.1 \times 10^8$ years.  This is a much shorter timescale than a Hubble time, and similar to that which \citet{vzee96} found for UGCA~020 ($\sim 8 \times 10^8$ years).  For most of the galaxies in \citet{vzee95} the free fall timescale was less than $\sim 7 \times 10^8$ years.  In contrast, \citet*{gio91} investigated the dynamically young "evolving" system \hi\ 1225+01 and found the free fall timescale to be of the order of $5 \times 10^9$ years, and the dynamical timescale to be even longer ($\sim 1 \times 10^{10}$ years).  This suggests that \esoq\ is not currently in the process of collapsing for the first time.  Instead, \citet{vzee96} suggest that this indicates the evolution process has been inhibited in some way.

This possibility is supported by the stellar population, which suggests that while some star formation is continuing today the majority of stars in \esoq\ are from past star formation events.  It certainly is not undergoing any vigorous star formation activity currently, and the past star formation rate suggests that it may never have undergone such a phase.  The stellar population and star formation rate  (present and past) are consistent with the galaxy having a low, more or less constant star formation rate since its formation.  As the galaxy shows no sign of disturbance, a slow, constant star formation rate is also consistent with the isolation of the galaxy.  Finally, the analysis of \hi\ gas surface density compared to the star formation threshold density derived from the \citeauthor{too64} stability criterion shows that gas density is probably too low for efficient star formation, although further analysis of \esoq's molecular gas content and of the validity of this criterion in dwarf irregulars is required.  \citet*{ver02} propose that, if angular momentum is conserved during gas contraction, low mass dark matter halos may form \citeauthor{too64} stable discs and therefore remain "dark" galaxies with little star formation.  \esoq\ is potentially such a galaxy, and if these dark galaxies do exist in sufficient numbers it could go some way to explain the current discrepancy between observed galaxy counts and the predictions from theoretical models such as $\Lambda$CDM.

\section{Conclusions}
\label{sec:conc}

In summary, we have examined the optical and radio properties of the gas-rich low surface brightness dwarf irregular galaxy \esoq.  We find that:
\begin{enumerate}
\item \esoq\ does have an unusually high \mlr\ of $22 \pm 4$\mls.  To our knowledge this is the highest \mlr\ for a stellar system to be confirmed by accurate measurement to date.
\item The \hi\ disk extends well beyond the stellar distribution, $6.4 \pm 0.4$ times the Holmberg radius ($r_{26.6}$ in the $B$ band) at a column density of $\sim 5.0 \times 10^{19}$ atoms\,cm$^{-2}$.
\item The galaxy is very isolated, with no neighbors identified out to 1 Mpc.
\item Optical colors suggest that the stellar population is older than is usually found in dwarf irregular galaxies.  Some shallow positive color gradients are observed between the inner and outer parts of the optical galaxy, suggesting that there are slightly more younger stars in the center of the galaxy.  However the large uncertainty caused by the high Galactic extinction prevents further analysis of the underlying stellar population.
\item \esoq\ has a relatively normal \hi\ content for a late-type dwarf galaxy, but is underluminous.
\item Star formation in \esoq\ is currently low, and may have remained relatively constant throughout the galaxies existence.
\item The \hi\ surface density of \esoq\ may be too low for the galaxy to efficiently form stars if the \citeauthor{too64} stability criterion determines the star formation threshold.
\end{enumerate}

\section*{Acknowledgments}
We are grateful for the assistance of Ken Freeman and Lister Staveley-Smith in this project, especially for their detailed comments on this paper before submission.  We would like to thank Agris Kalnajs for his assistance with the star formation threshold analysis, and Frank Briggs for his suggestions for \hi\ flux density calculation.  We would also like to thank the anonymous referee for their useful comments.  The 2.3-meter Telescope is run by the Australian National University as part of Research School of Astronomy and Astrophysics.  The Australia Telescope Compact Array and the Parkes Radio Telescope are part of the Australia Telescope which is funded by the Commonwealth of Australia for operation as a National Facility managed by CSIRO.  This research has made use of the NASA/IPAC Extragalactic Database (NED) which is operated by the Jet Propulsion Laboratory, California Institute of Technology, under contract with the National Aeronautics and Space Administration.  The Digitized Sky Survey (DSS) was produced at the Space Telescope Science Institute under U.S. Government grant NAG W-2166, based on photographic data obtained using the UK Schmidt Telescope.

\clearpage

\begin{deluxetable}{llll} 
\tabletypesize{\scriptsize}
\tablecaption{Properties of \esoq.  
\label{tab:prop}}
\tablewidth{0pt}
\tablehead{ \colhead{Property} & \colhead{Previous Value}   & \colhead{Reference} & \colhead{This Paper} }
\startdata
Names                  & \esoq\                            & \citet{lau82}  \\
                       & HIPASS J1057--48                  & BGC            \\
                       & PGC~490287                        & LEDA           \\
                       & KKs40                             & \citet{kara00} \\
Center position        & $10^{\rm h}\,57^{\rm m}$\,29\fs37 & \citet{lau82} & Optical: $10^{\rm h}\,57^{\rm m}$\,29\fs9 \\
~$\alpha,\delta$(J2000)& --48\degr\,10\arcmin\,40\farcs1~($\pm$8\arcsec)   & & --48\degr\,10\arcmin\,43\arcsec~($\pm$2\arcsec) \\
                       & & & \hi: $10^{\rm h}\,57^{\rm m}$\,30\fs0 \\
                       & & & --48\degr\,10\arcmin\,47\arcsec~($\pm$10\arcsec) \\
Galactic coordinates $l, b$        & 284\fdg1, 10\fdg5                   &                \\
systemic velocity, $v_{\rm sys}$   & $598 \pm 1$\kms       & BGC & $597 \pm 1$\kms \\
50\% velocity width, $w_{50}$      & $ 67 \pm 2$\kms       & BGC &  \\
20\% velocity width, $w_{20}$      & $ 83 \pm 3$\kms       & BGC &  \\
Local group velocity, $v_{\rm LG}$ & $312 \pm 1$\kms       & BGC & $311 \pm 1$\kms \\
Distance, $D$                      & 4.2 Mpc   & \\
apparent $R$ magnitude, $m_{\rm R}$ & $15.02 \pm 0.09$ mag & \citet{lau89} & $14.38 \pm 0.11$ mag \\
apparent $B$ magnitude, $m_{\rm B}$ & $16.03 \pm 0.09$ mag & \citet{lau89} & $16.13 \pm 0.07$ mag \\
 & $16.43 \pm 0.41$ mag & LEDA \\
extinction, \AB                    & $0.95 \pm 0.15$ mag & \citet{sch98} \\
                                   & 0.84 mag          & \citet{bur82} \\
absolute $B$ magnitude, $M_{B,0}$    & $-12.64 \pm 0.4$ mag      &  & $-12.9 \pm 0.2$ mag \\
\hi\ flux density, \FHI\    & $104.4 \pm 11.5$\jks & BGC  & $122 \pm 4$\jks \\
\hi\ mass, \MHI\            & $(4.3 \pm 0.5) \times 10^{8}$\Msun  & BGC  & $(5.1 \pm 0.2) \times 10^{8}$\Msun \\
\mlr                               & $24 \pm 12$\mls   & BGC $+$ LEDA  & $22 \pm 4$\mls \\
\enddata
\end{deluxetable}

\begin{deluxetable}{lccccccccc} 
\tabletypesize{\scriptsize}
\tablecaption{{\em BVRI} Photometry Results for \esoq.  
\label{tab:mag}}
\tablewidth{0pt}
\tablehead{ \colhead{Band} & \colhead{$m_{T}$ \tablenotemark{a}} & \colhead{$m_{\mbox{S\'ersic}}$ \tablenotemark{a}} & \colhead{$\mu_{0}$ \tablenotemark{a}} & \colhead{$\langle \mu \rangle _{eff}$ \tablenotemark{a}} & \colhead{$r_{eff}$} & \colhead{$r_{26.6}$} & \colhead{\AG} \\

\colhead{} & \colhead{(mag)} & \colhead{(mag)} & \colhead{(mag arcsec$^{-2}$)} & \colhead{(mag arcsec$^{-2}$)} & \colhead{(arcsec)} & \colhead{(arcsec)} & \colhead{(mag)} \\

\colhead{(1)} & \colhead{(2)} & \colhead{(3)} & \colhead{(4)} & \colhead{(5)} & \colhead{(6)} & \colhead{(7)} & \colhead{(8)} }
\startdata
{\em B} & $16.13 \pm 0.07$ & $16.03 \pm 0.07$ & $24.97 \pm 0.03$ & $25.48 \pm 0.02$ & $29.7 \pm 0.6$ & $57.6 \pm 0.6$ & $0.95 \pm 0.15$ \\
{\em V} & $14.89 \pm 0.06$ & $14.83 \pm 0.06$ & $23.65 \pm 0.03$ & $24.14 \pm 0.02$ & $28.3 \pm 0.6$ & $65 \pm 2$ & $0.73 \pm 0.12$ \\
{\em R} & $14.38 \pm 0.11$ & $14.31 \pm 0.11$ & $23.16 \pm 0.02$ & $23.64 \pm 0.03$ & $28.4 \pm 0.6$ & $68 \pm 2$ & $0.59 \pm 0.09$ \\
{\em I} & $13.76 \pm 0.13$ & $13.70 \pm 0.13$ & $22.91 \pm 0.04$ & $23.40 \pm 0.04$ & $33.9 \pm 0.6$ & ~$85 \pm 10$ & $0.43 \pm 0.07$ \\
\enddata
\tablenotetext{a}{Galactic extinction not applied.}
\end{deluxetable}

\begin{deluxetable}{lrrll} 
\tabletypesize{\scriptsize}
\tablecaption{Comparison of \esoq\ Properties with those of other Late-Type Dwarf Galaxies.  \label{tab:typical}}
\tablewidth{0pt}
\tablehead{ \colhead{Property} & \colhead{\esoq} & \colhead{Late-Type} & \colhead{Reference} & \colhead{Sample} \\

\colhead{} & \colhead{(This paper)} & \colhead{Dwarf Galaxies} & \colhead{} & \colhead{} \\

\colhead{(1)} & \colhead{(2)} & \colhead{(3)} & \colhead{(4)} & \colhead{(5)} }
\startdata
\MB\ (mag)                      & $-12.9 \pm 0.2$  & $-16.8$ to $-18.6$ & \citet{rob94} & Sm/Im (quartile range) \\
 &  & $-11.53$ to $-18.28$ & \citet{lee03} & Field dI's \\
 &  & $-12.8$ to $-17.6$ & \citet{sti02a} & Im \& Sm \\
$\mu_{B,0}$ (mag arcsec$^{-2}$) & $24.97 \pm 0.03$ & $24.33$ to $20.31$ & \citet{bar01} & Sm/Im/BCD \\
\LB\ ($\times 10^{7}$ \Lsun)    & $2.3 \pm 0.4$    & $80$ to $410$      & \citet{rob94} & Sm/Im (quartile range) \\
\vmax\ (\kms)            & $51 \pm 8$       & $33$ to $89$ & \citet{sti02b} & Im \& Sm \\
\MHI\ ($\times 10^{8}$ \Msun)   & $5.1 \pm 0.2$    & $4.3$ to $26.9$    & \citet{rob94} & Sm/Im (quartile range) \\
 &  & $0.098$ to $17.4$ & \citet{lee03} & Field dI's \\
 &  & $0.04$ to $16.56$ & \citet{sti02a} & Im \& Sm \\
\mlr\ (\mls)                    & $22 \pm 4$       & $0.44$ to $1.32$   & \citet{rob94} & Sm/Im (quartile range) \\
 &  & $0.15$ to $4.2$ & \citet{lee03} & Field dI's \\
 &  & $0.1$ to $2.8$ & \citet{sti02a} & Im \& Sm \\
\Mtot\ ($\times 10^{9}$ \Msun)  & $4.5 \pm 1.6$    & $4$ to $18$        & \citet{rob94} & Sm/Im (quartile range) \\
\mtmr                           & $0.11 \pm 0.04$  & $0.09$ to $0.24$   & \citet{rob94} & Sm/Im (quartile range) \\
\tmlr\ (\mls)                   & $200 \pm 110$    & $3.1$ to $7.9$     & \citet{rob94} & Sm/Im (quartile range) \\
$(B - V)_{0}$ (mag)             & $1.02 \pm 0.21$  & $0.35$ to $0.51$   & \citet{rob94} & Sm/Im (quartile range) \\
 &  & $-0.18$ to $0.73$ & \citet{sti02a} & Im \& Sm \\
 &  & $0.07$ to $0.71$ & \citet{bar01} & Sm/Im/BCD  \\
$(B - R)_{0}$ (mag)             & $1.39 \pm 0.22$  & $0.61$ to $1.24$   & \citet{par02} & Sd/Sm/Im/BCD \\
 &  & $0.34$ to $1.24$ & \citet{bar01} & Sm/Im/BCD  \\
$(B - I)_{0}$ (mag)             & $1.85 \pm 0.22$  & $-$              &  &  \\
$(V - R)_{0}$ (mag)             & $0.37 \pm 0.20$  & $0.10$ to $0.71$ & \citet{bar01} & Sm/Im/BCD  \\
$(V - I)_{0}$ (mag)             & $0.83 \pm 0.20$  & $-$              &  &  \\
$(R - I)_{0}$ (mag)             & $0.46 \pm 0.20$  & $-$              &  &  \\
\enddata
\end{deluxetable}

\begin{deluxetable}{lrr} 
\tabletypesize{\scriptsize}
\tablecaption{Comparison of Distance Dependent Properties for \esoq\ at Different Adopted Distances.  \label{tab:distdep}}
\tablewidth{0pt}
\tablehead{ \colhead{Property} & \colhead{Hubble Flow} & \colhead{Baryonic Tully-Fisher} \\

\colhead{} & \colhead{{\it D} = 4.2 Mpc} & \colhead{{\it D} = 2.6 Mpc} }
\startdata
\MB\ (mag)                     & $-12.9 \pm 0.2$  & $-11.9 \pm 0.2$ \\
\LB\ ($\times 10^{7}$ \Lsun)   & $2.3 \pm 0.4$    & $0.9 \pm 0.1$ \\
\MHI\ ($\times 10^{8}$ \Msun)  & $5.1 \pm 0.2$    & $1.95 \pm 0.06$ \\
\Mtot\ ($\times 10^{9}$ \Msun) & $4.5 \pm 1.6$    & $2.8 \pm 1.0$ \\
\mtmr                          & $0.11 \pm 0.04$  & $0.07 \pm 0.02$ \\
\tmlr\ (\mls)                  & $200 \pm 110$    & $320 \pm 150$ \\
\enddata
\end{deluxetable}

\clearpage

\begin{figure*} 
\begin{tabular}{cc}
 \mbox{\psfig{figure=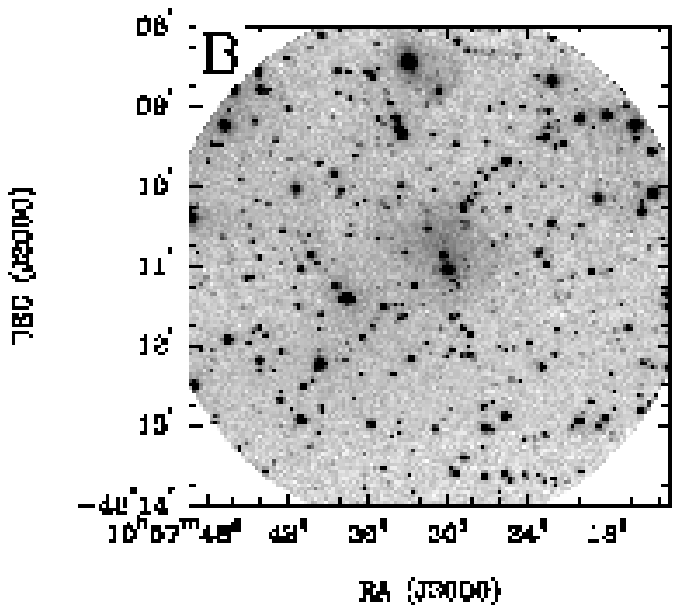,width=8cm}} & 
 \mbox{\psfig{figure=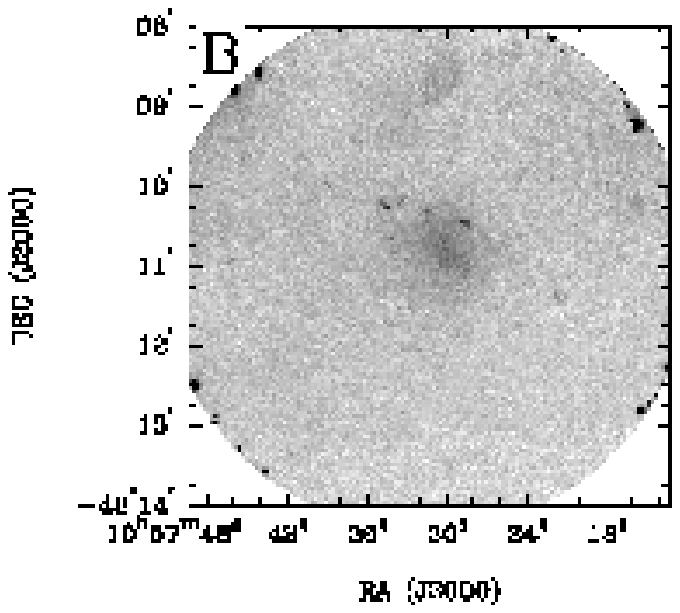,width=8cm}} \\
 \mbox{\psfig{figure=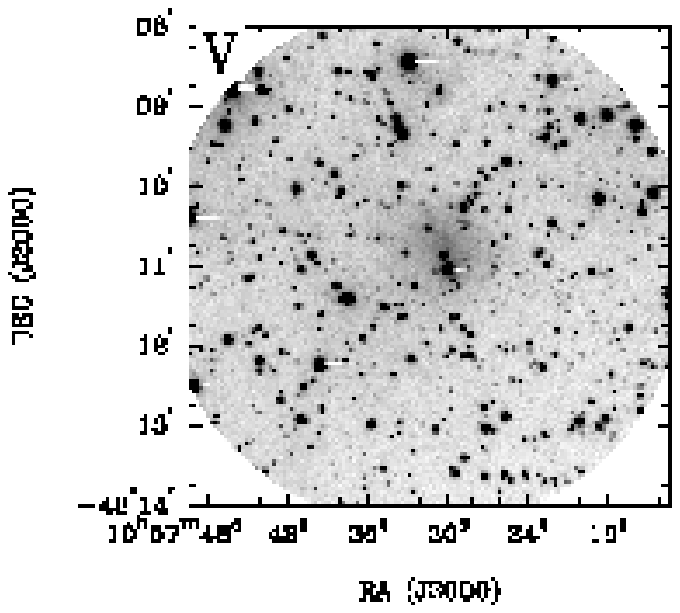,width=8cm}} & 
 \mbox{\psfig{figure=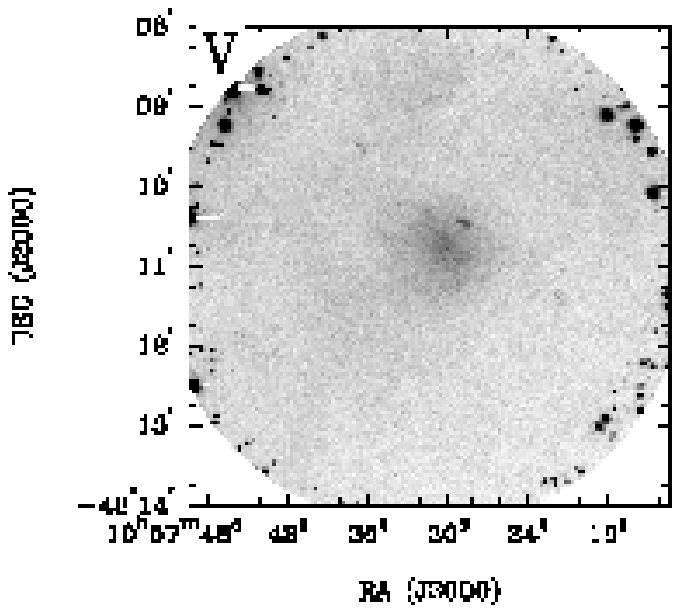,width=8cm}} \\
\end{tabular}
\caption{Deep {\em BVRI} images of \esoq\ before and after subtraction of the foreground stars.  East is to the left and North is up for all images in this paper.
\label{fig:optimage}}
\end{figure*}
\setcounter{figure}{1}
\begin{figure*} 
\begin{tabular}{cc}
 \mbox{\psfig{figure=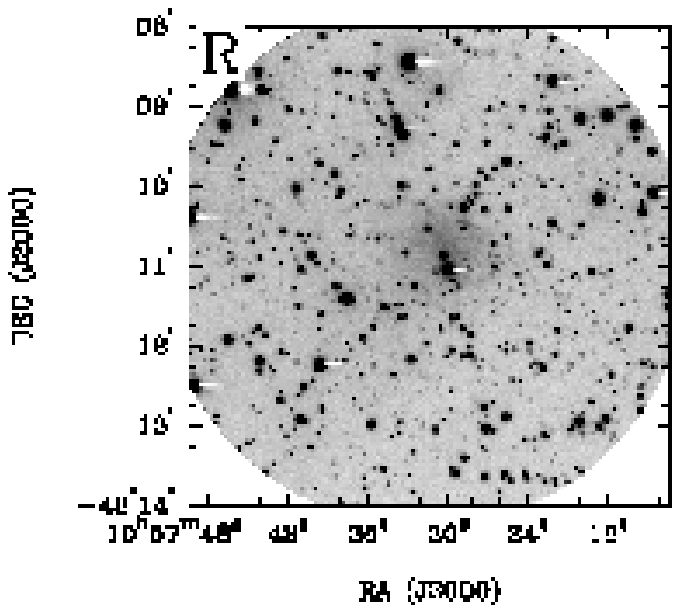,width=8cm}} & 
 \mbox{\psfig{figure=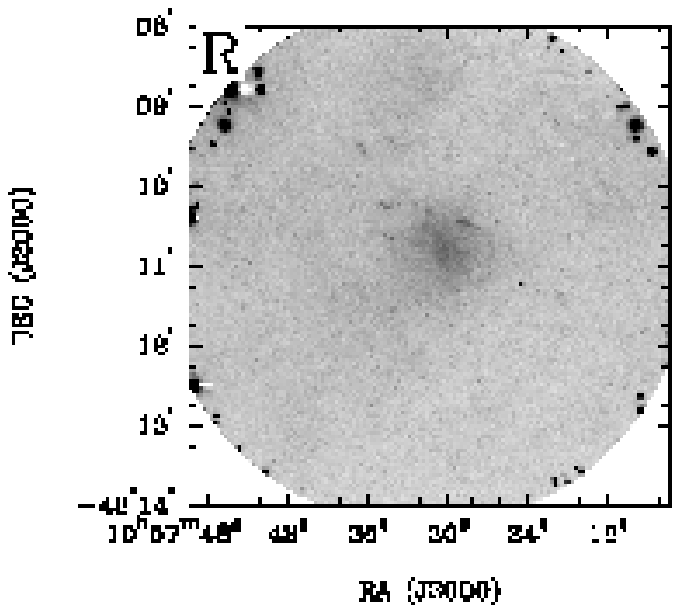,width=8cm}} \\
 \mbox{\psfig{figure=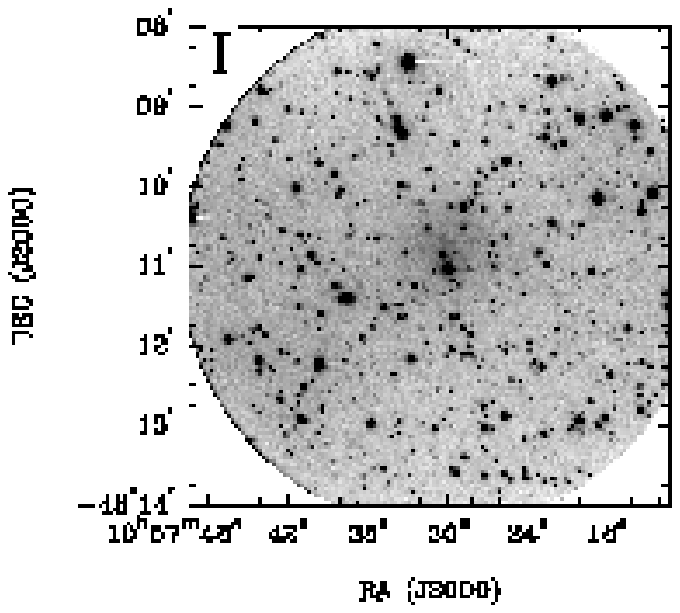,width=8cm}} & 
 \mbox{\psfig{figure=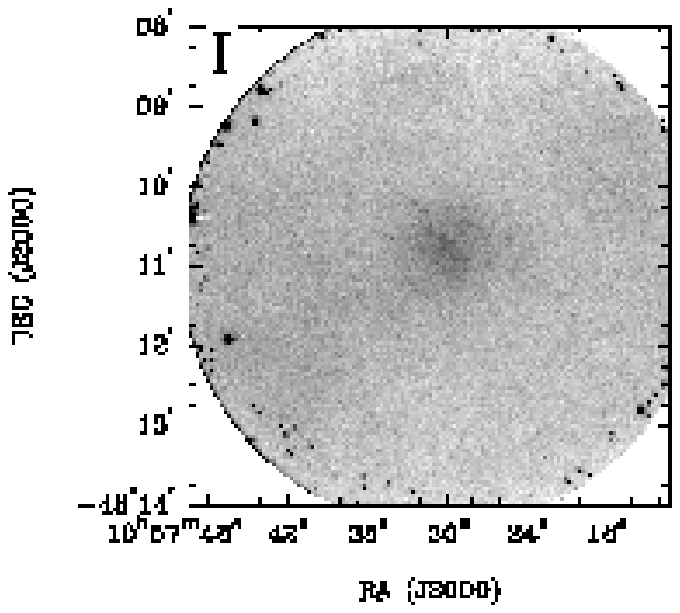,width=8cm}} \\
\end{tabular}
\caption{continued}
\end{figure*}

\begin{figure} 
 \centering{\psfig{figure=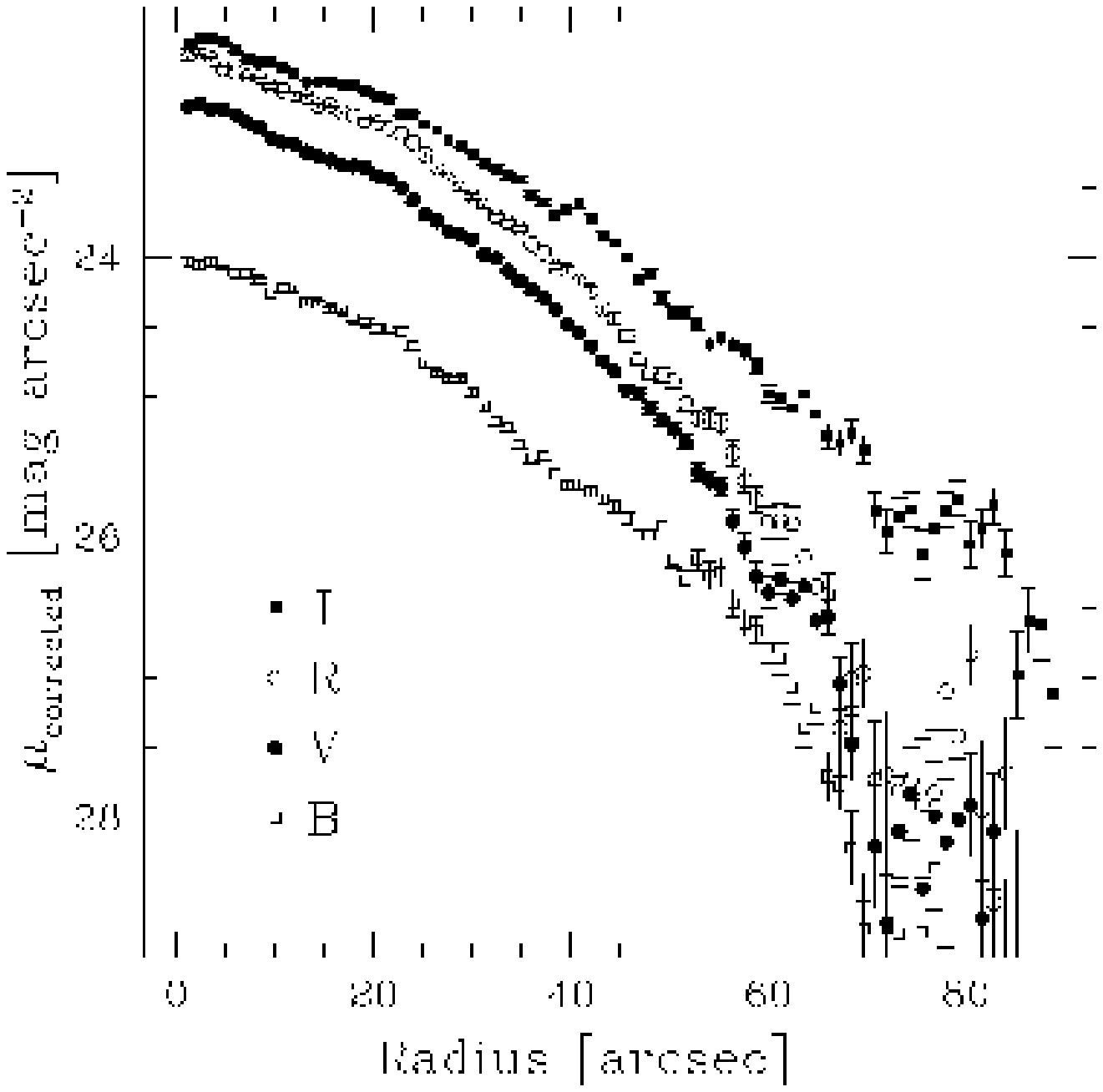,width=8cm}} 
\caption{{\em BVRI} surface brightness profiles of \esoq, corrected for Galactic extinction.
\label{fig:optprofile}}
\end{figure}

\begin{figure} 
 \centering{\psfig{figure=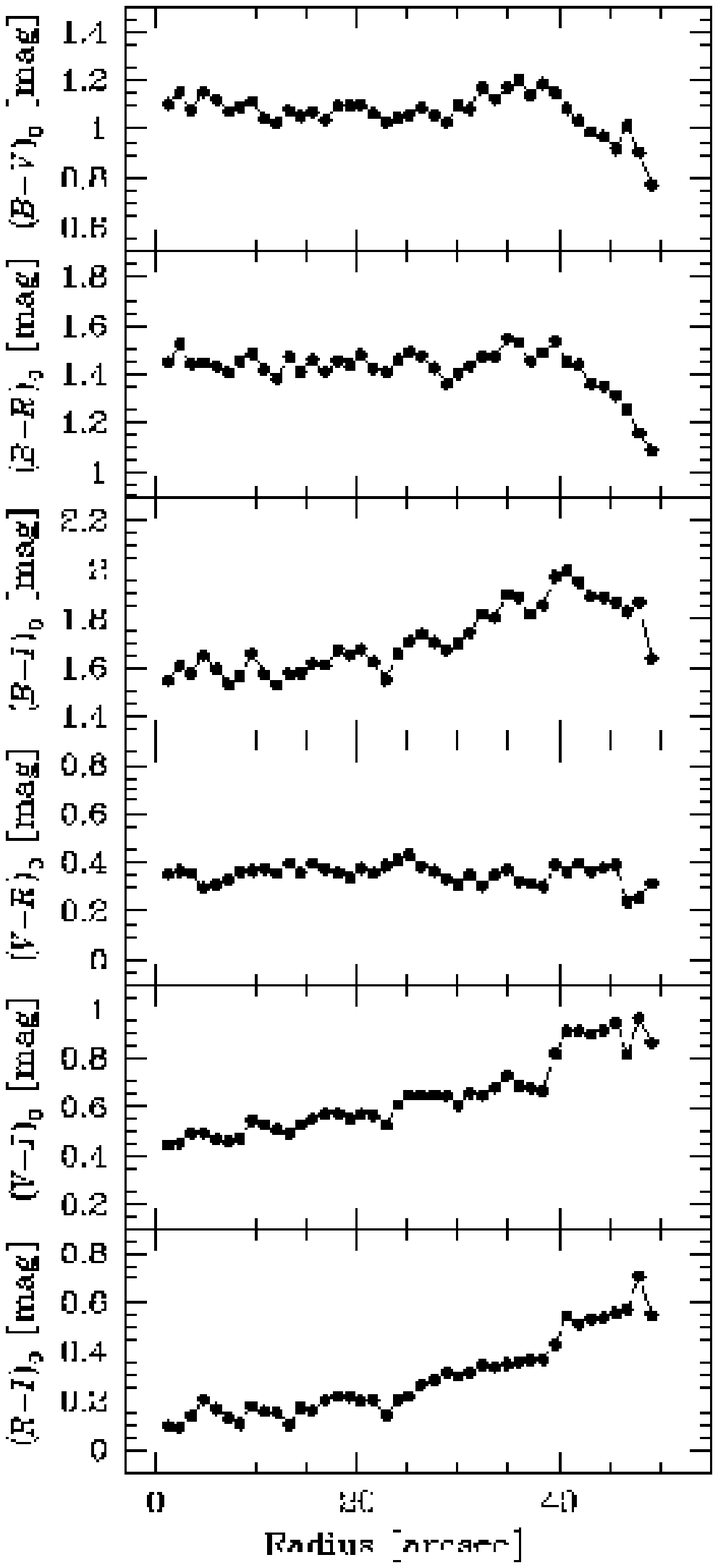,width=8cm}} 
\caption{Color Profiles for \esoq, corrected for Galactic extinction.
\label{fig:colourprofile}}
\end{figure}

\begin{figure*} 
\centering{\psfig{figure=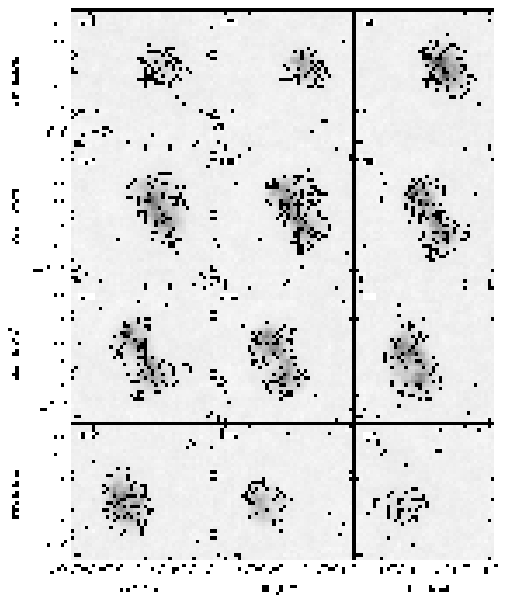,width=15.5cm}}
\caption{\hi\ channel maps of \esoq. For display purposes we show channels
    smoothed to a velocity resolution of 8\kms. The contour levels are 
    $\pm$3.15 ($\sim3\sigma$), 8.4, 15.75, 26.25, 42, and 63 mJy\,beam$^{-1}$. 
    The greyscale ranges from --6 to 95 mJy\,beam$^{-1}$.  Negative contours are
    marked with dashed lines.
    The cross marks the dynamical center of the galaxy. The synthesized
    beam is displayed in the bottom left corner of each panel. The central
    heliocentric velocity for each panel is displayed in the top left corner.
\label{fig:hichannel}}
\end{figure*}

\begin{figure} 
 \centering{\psfig{figure=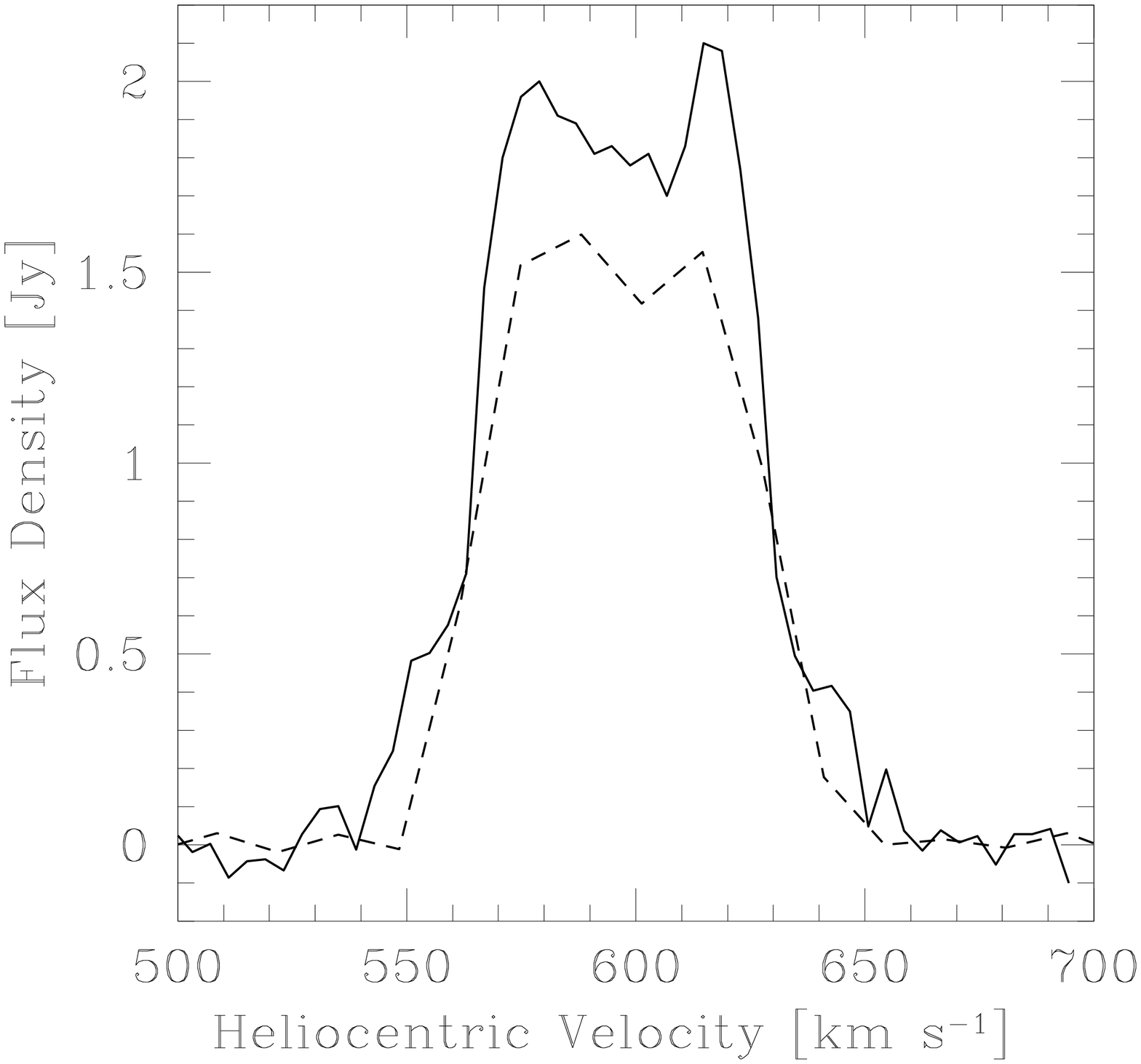,width=8cm}} 
\caption{Global \hi\ spectra of \esoq\ as obtained from HIPASS 
  (dashed lines) and the ATCA (solid line).
\label{fig:hispectra}}
\end{figure}

\begin{figure*} 
\begin{tabular}{ll}
 (a) & (b) \\
 \mbox{\psfig{figure=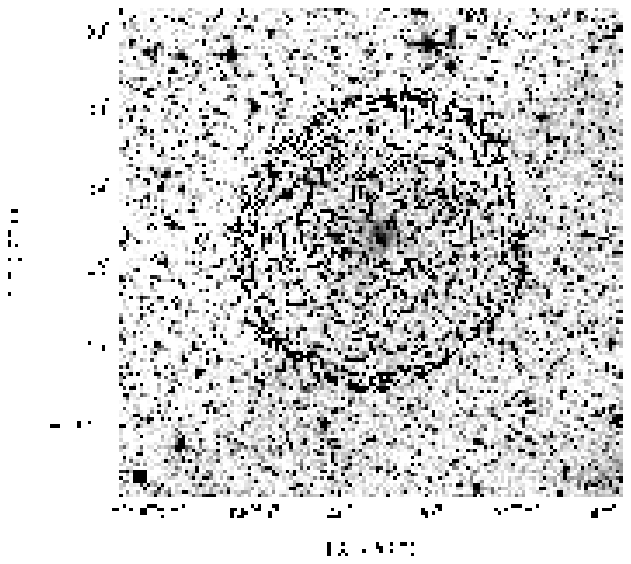,height=6.5cm}} & 
 \mbox{\psfig{figure=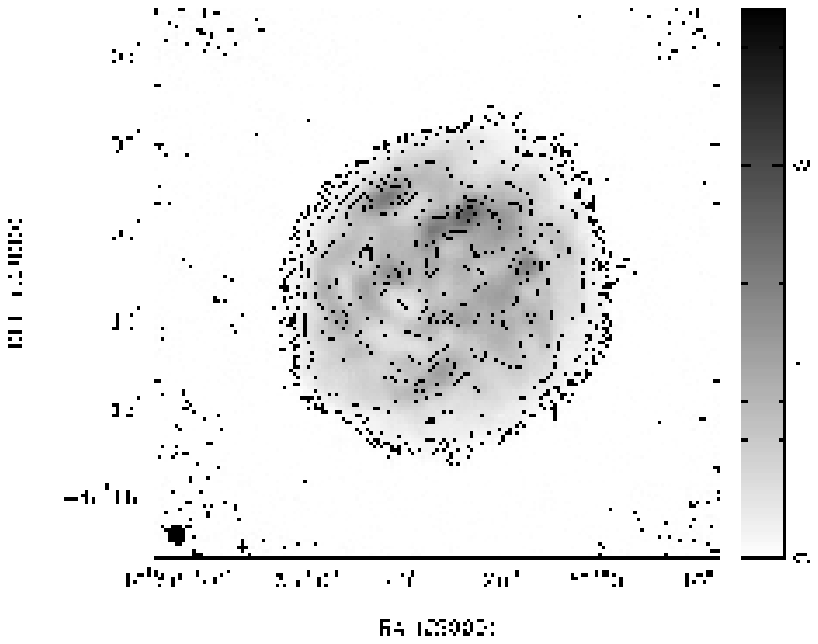,height=6.5cm}} \\
 (c) & (d) \\
 \mbox{\psfig{figure=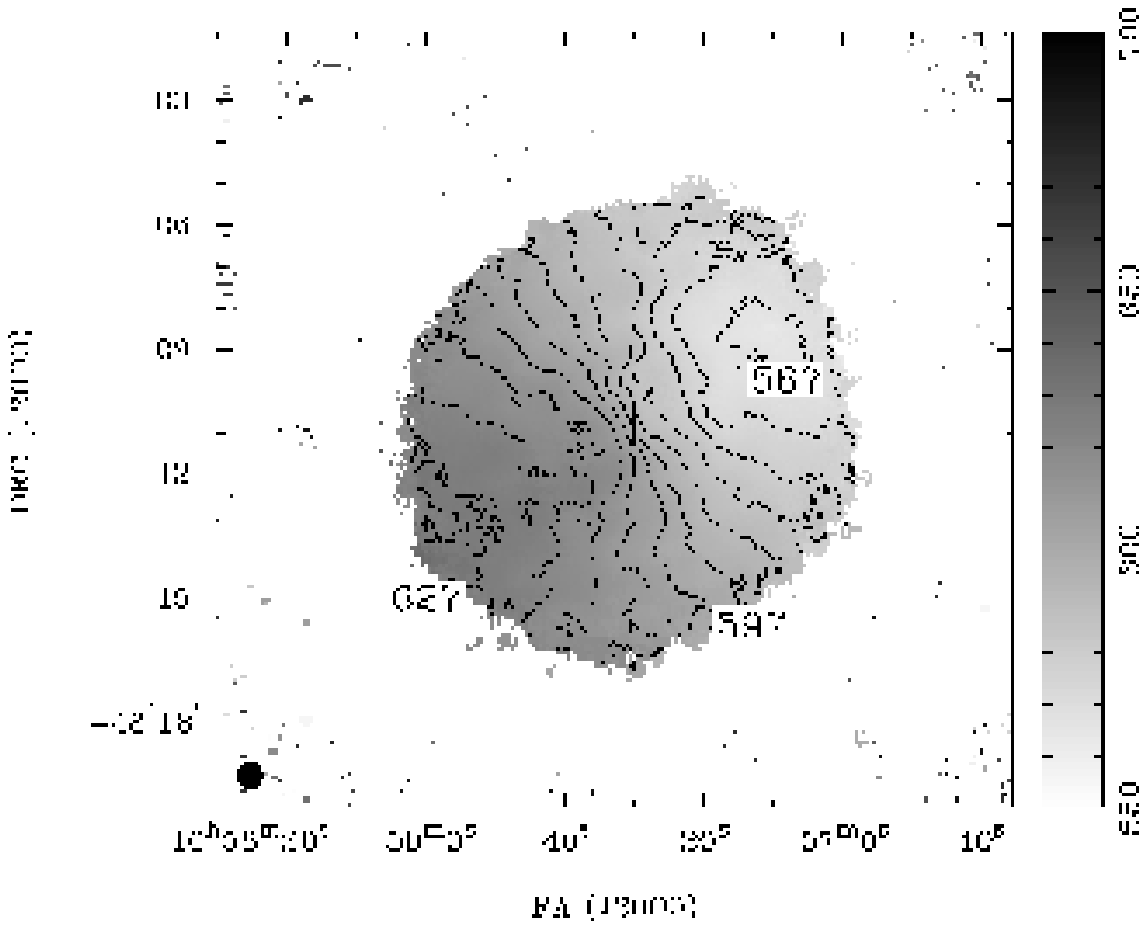,height=6.5cm}} & 
 \mbox{\psfig{figure=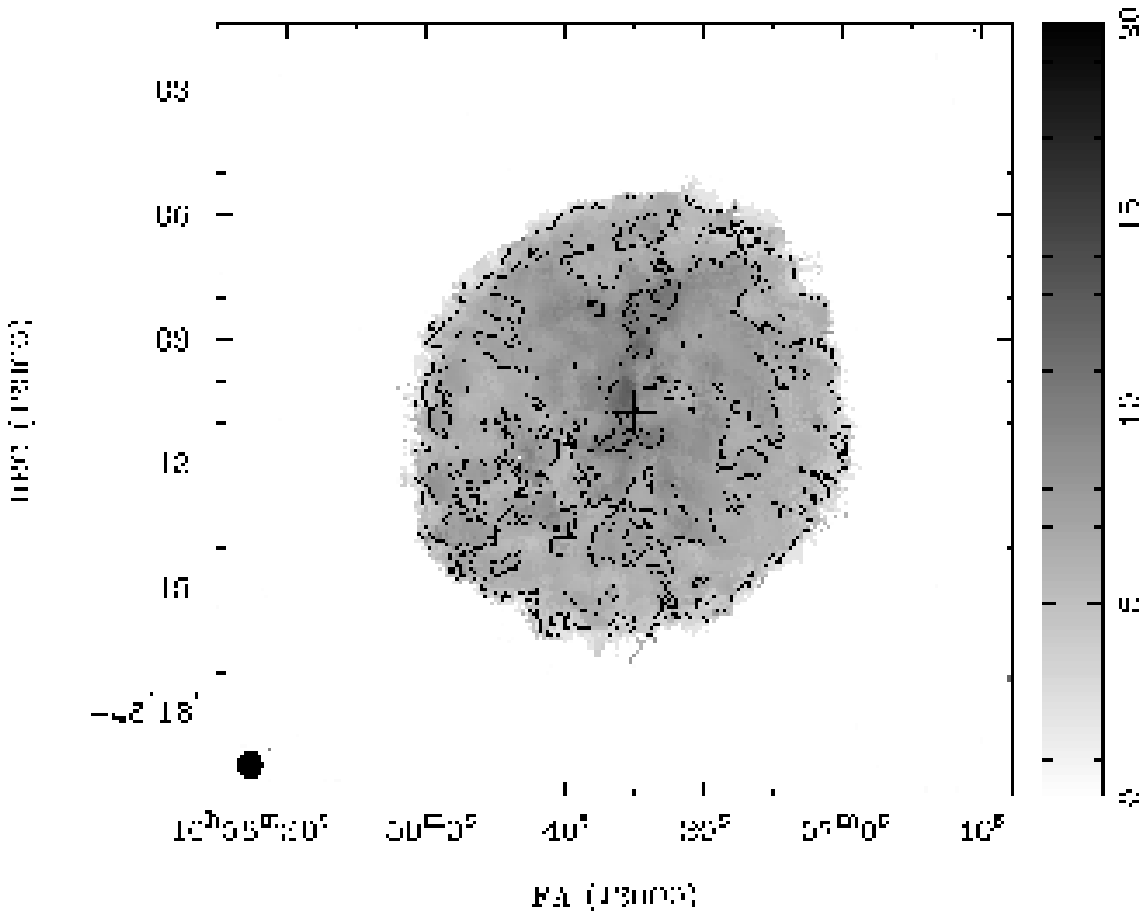,height=6.5cm}} \\
\end{tabular}
\caption{\hi\ moment maps of \esoq. 
    (a) Integrated \hi\ intensity distribution overlaid onto an optical 
        DSS\,II {\em R} band image. The contour levels are 0.073, 0.145, 
        0.29, 0.58, 0.87, 1.16 and 1.57 Jy\,beam$^{-1}$\kms\ (which corresponds 
	to column densities of 6, 12, 24, 48, 72, 96 and $130 \times
	10^{19}$ atoms cm$^{-2}$, respectively),
    (b) as in (a) but overlaid onto itself, and with the additional contours 
        of $\pm$0.036 Jy\,beam$^{-1}$\kms\ ($\pm$3 $\times
	10^{19}$ atoms cm$^{-2}$).  Negative contours are dashed,
    (c) Mean \hi\ velocity field. The contours levels range from 
        567 to 627\kms\ in steps of 6\kms,
    (d) \hi\ velocity dispersion. The contours levels are 4, 7, and 10\kms.
    The cross in (b), (c), and (d) marks the dynamical center of the galaxy.  
    The synthesized beam is displayed in the bottom left corner of each panel. 
\label{fig:mom}}
\end{figure*}

\begin{figure} 
\begin{tabular}{cc}
 \mbox{\psfig{figure=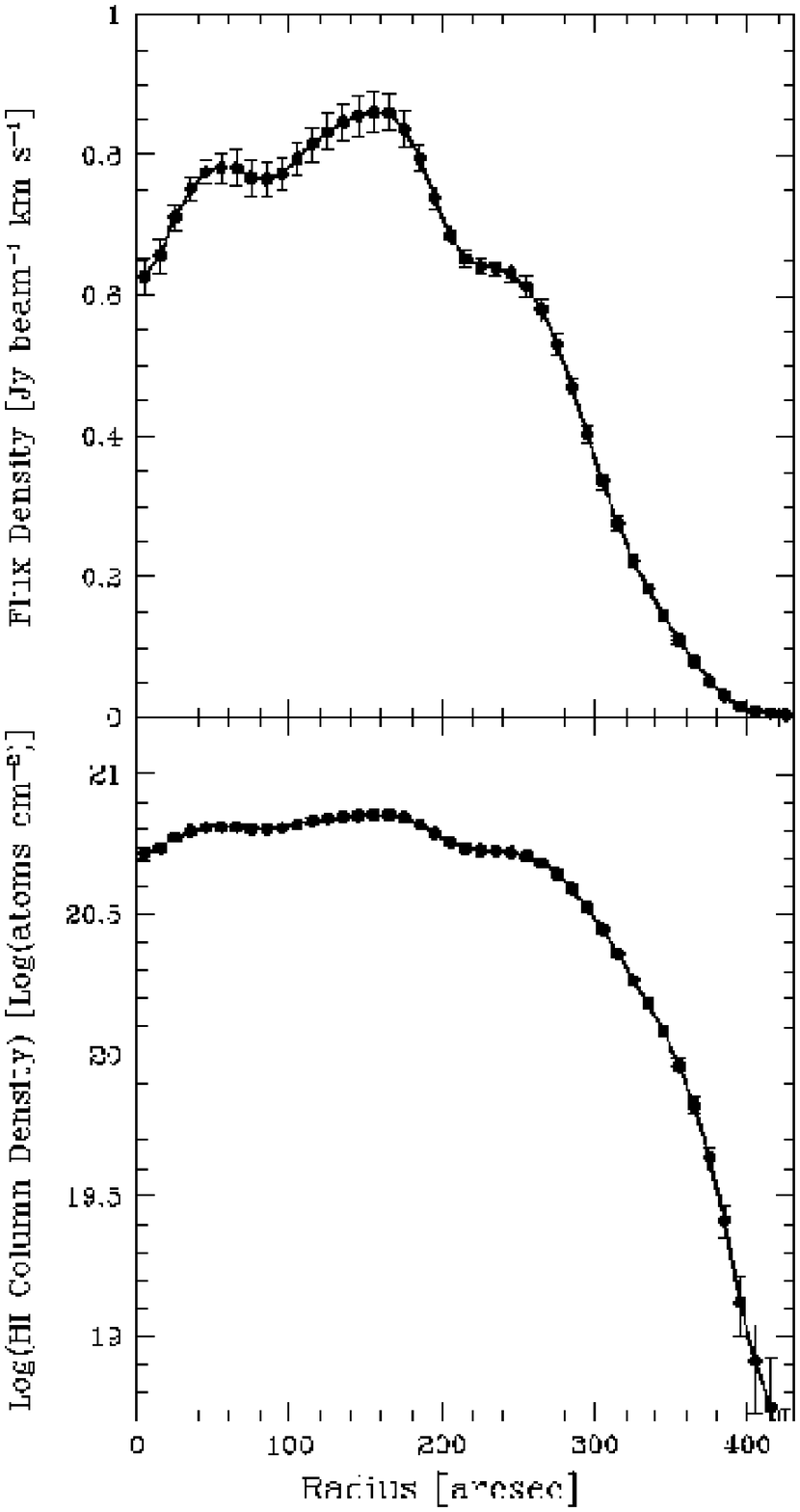,width=8cm}}
\end{tabular} 
\caption{Azimuthally averaged \hi\ radial profiles of \esoq\ in (a) flux 
  density, and (b) column density.  The profile was measured in 10\arcsec\ 
  steps, and corrected for an inclination of 36\degr.  Error bars reflect 
  Poisson statistical uncertainties.
\label{fig:hirad}}
\end{figure}

\begin{figure} 
 \centering{\psfig{figure=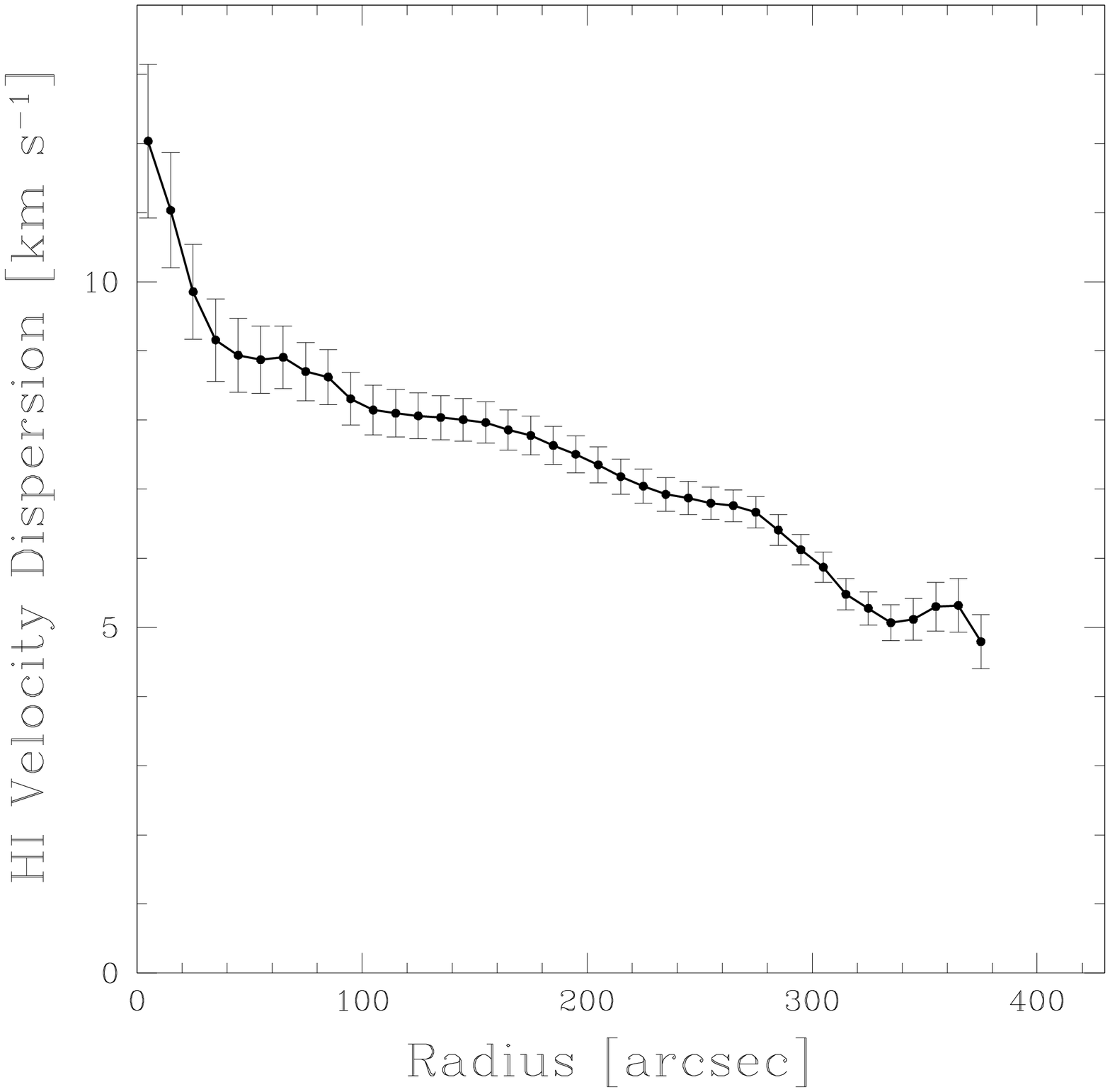,width=8cm}} 
\caption{Azimuthally averaged \hi\ velocity dispersion radial profile of 
  \esoq.  The profile was measured in 10\arcsec\ steps, and corrected for 
  an inclination of 36\degr.  Error bars reflect Poisson statistical 
  uncertainties.
\label{fig:veldisp}}
\end{figure}

\begin{figure} 
 \centering{\psfig{figure=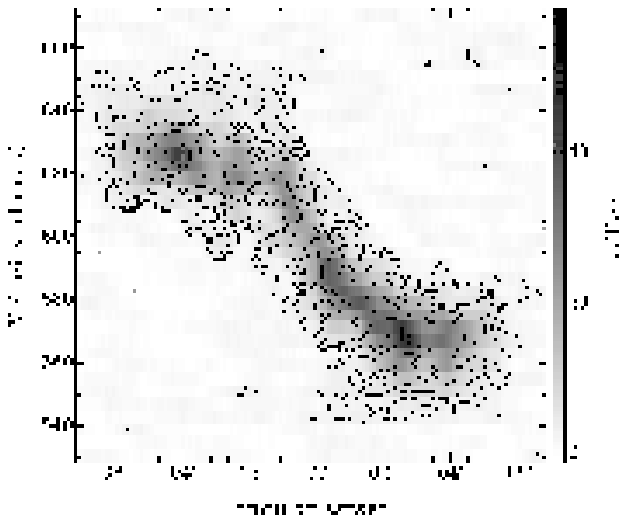,width=8cm}} 
\caption{\hi\ position velocity diagram along the major axis of \esoq\ at a 
  position  angle of 119\degr.  The slice is averaged over a width of 
  $\sim$40\arcsec\ (slightly larger than the beam size of the \hi\ 
  observations).  Contour levels are $\pm$2.0, 2.8, 4.0, 5.7, 8.0, 11.3, 16.0, 
  22.6, 32.0, and 45.3 mJy\,beam$^{-1}$.  Negative contours are in light grey.
\label{fig:pvd}}
\end{figure}

\begin{figure} 
 \centering{\psfig{figure=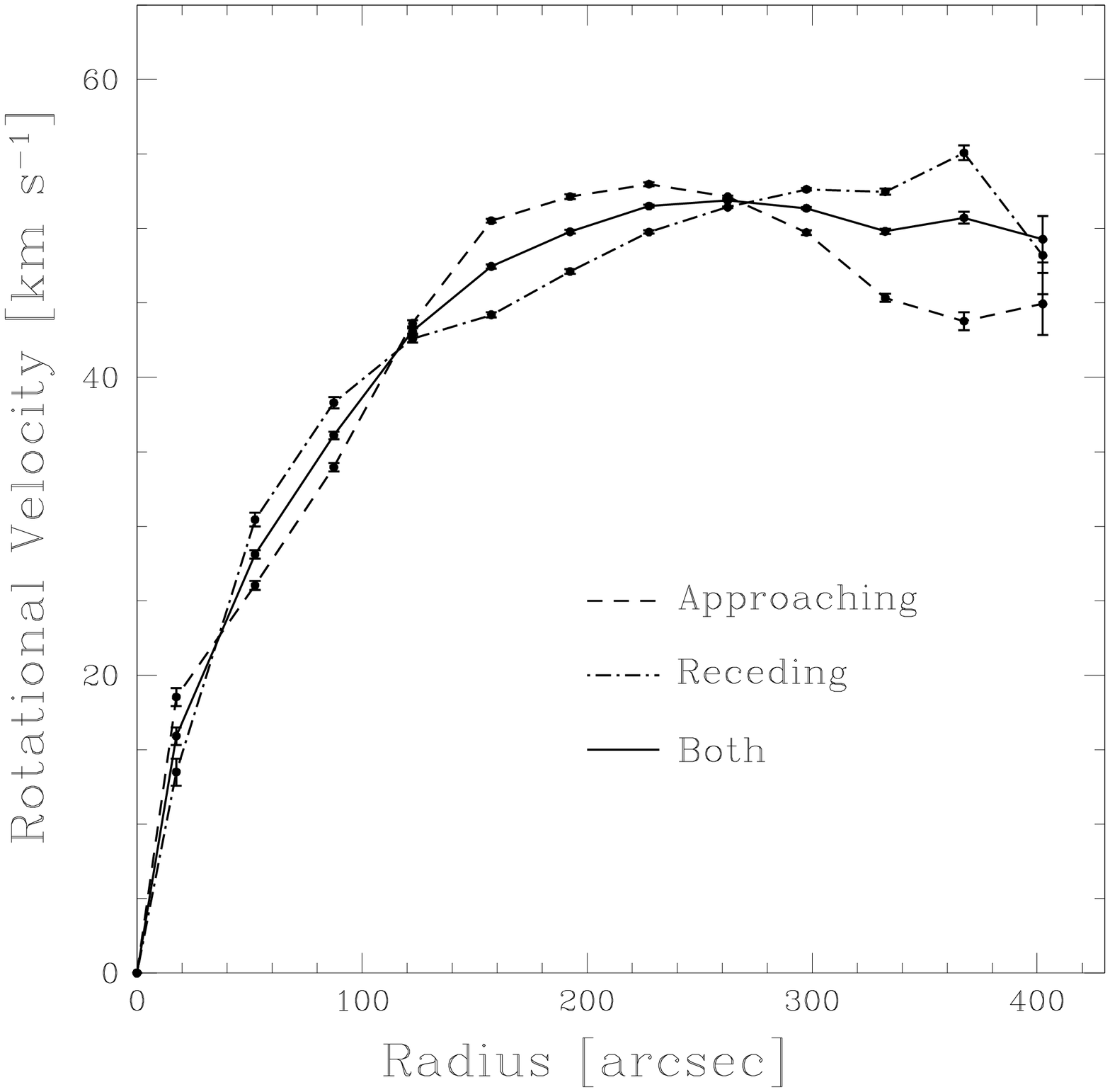,width=8cm}} 
\caption{Rotation curve of \esoq\ as derived from the mean \hi\ velocity
  field. The three curves correspond to the approaching side (dashed line), 
  receding side (dot-dashed line), and both sides (solid line). For the final 
  {\sc rocur} fit the inclination angle was fixed at 36\degr, and the position 
  angle at 119\degr. For details see Section~\ref{sec:velo}.
\label{fig:rotc}}
\end{figure}

\begin{figure} 
 \centering{\psfig{figure=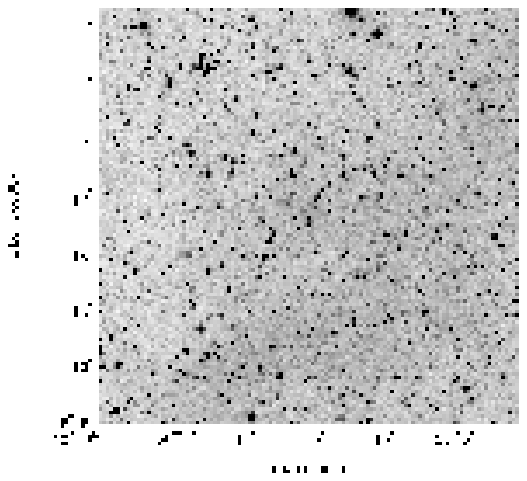,width=8cm}} 
\caption{20-cm radio continuum map overlaid on the SHS H$\alpha$ image of the 
  field around \esoq.
\label{fig:20cmha}}
\end{figure}

\begin{figure} 
 \centering{\psfig{figure=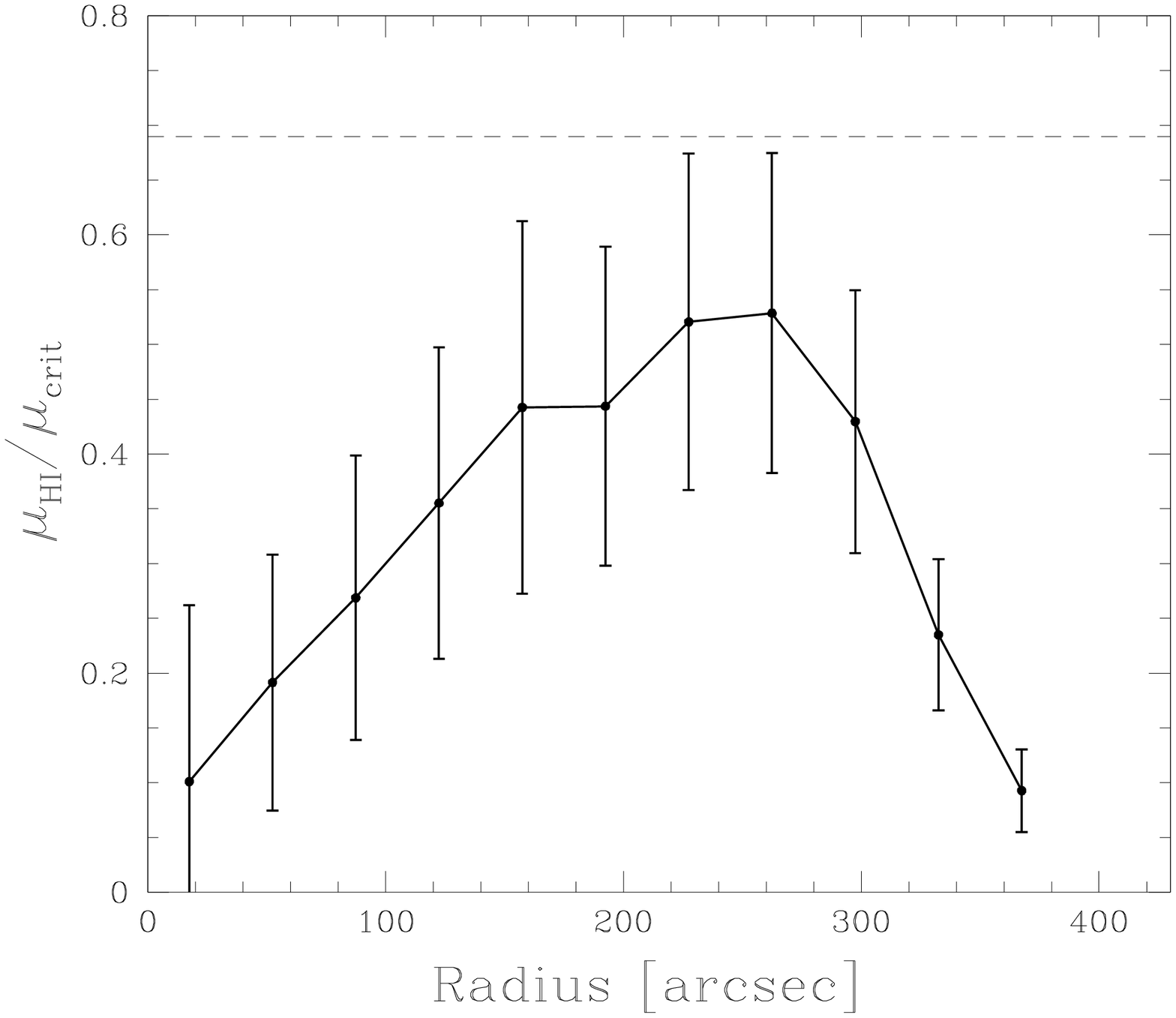,width=8cm}} 
\caption{The ratio of \hi\ surface density to the critical surface density (as 
  predicted from the \citet{too64} thin disk gravitational stability Q 
  criterion for thin disks) as a function of radius for \esoq.  The dashed
  line shows $\alpha_{Q} = 0.69$, the median value from \citet{mar01} above 
  which the gas density is high enough for large scale star formation (see 
  Section~\ref{sec:dis-toomre}).  Error bars reflect the uncertainties in the 
  \hi\ radial profile, \hi\ velocity dispersion profile and rotation curve.
\label{fig:toomreq}}
\end{figure}

\end{document}